\documentclass[letterpaper,journal]{IEEEtran}
\usepackage{amsmath,amsfonts}

\usepackage{array}
\usepackage{stfloats}
\usepackage{url}
\usepackage{verbatim}
\usepackage{graphicx}
\usepackage{cite}
\usepackage{color}
\usepackage{booktabs}
\usepackage{amssymb}
\usepackage{makecell}
\usepackage{listings}
\usepackage{fontawesome} 
\usepackage[table,dvipsnames]{xcolor}
\usepackage{hyperref}
\usepackage{multirow}
\hypersetup{
    colorlinks=true,
    linkcolor=blue,
    urlcolor=blue,  
    citecolor=blue}
\usepackage{colortbl}
\definecolor{bg}{HTML}{e0f1ff}
\newcommand{\classswatch}[1]{\cellcolor[HTML]{#1}\hspace{0.78cm}}

\begin{document}

\title{Shared-Pattern Collaborative Network for Multi-Source Remote Sensing Data Classification}

\author{
Chenfang Yang,
Chuanzheng Gong,
Feng~Gao, \IEEEmembership{Member,~IEEE,} \\
Junyu~Dong, \IEEEmembership{Member,~IEEE,} and 
Qian~Du, \IEEEmembership{Fellow,~IEEE}

\thanks{This work was supported in part by the Natural Science Foundation of Shandong Province under Grant ZR2024MF020, in part by the Key R \& D Program of Shandong Province under Grant 2025CXPT185 and in part by the National Natural Science Foundation of China under Grant 42676229.
\textit{(Corresponding author: Feng Gao)}}
\thanks{Chenfang Yang, Chuanzheng Gong, Feng Gao, and Junyu Dong are with the State Key Laboratory of Physical Oceanography, Ocean University of China, Qingdao 266100, China. (e-mail: gaofeng@ouc.edu.cn)}
\thanks{Qian Du is with the Department of Electrical and Computer Engineering, Mississippi State University, Starkville, MS 39762 USA.}}

\markboth{IEEE Journal of Selected Topics in Applied Earth Observations and Remote Sensing}%
{Yang \MakeLowercase{\textit{et al.}}: Shared-Pattern Collaborative Network for Multi-Source Remote Sensing Data Classification}

\maketitle

\begin{abstract}
Multi-source remote sensing data classification has received growing attention recently. However, existing methods still struggle to jointly model long-range contextual dependencies and shared spatial patterns across heterogeneous modalities. To address this issue, we propose a Shared-Pattern Collaborative Network (SPCNet) for HSI and SAR/LiDAR joint classification. To enhance cross-modal collaborative learning, \textit{Shared-Pattern Collaborative Block (SPCB)} is designed by introducing a shared large-kernel operator and a dynamic convolution branch, enabling the network to capture common spatial structures and content-adaptive local features. Furthermore, \textit{Region-Adaptive Modality Allocator (RAMA)} is developed to assign spatially varying modality weights for reliable feature fusion. Experiments on Houston2013, Augsburg, and Houston2018 datasets demonstrate that SPCNet achieves superior classification performance over several state-of-the-art methods. The code will be made publicly available at \url{https://github.com/oucailab/SPCNet}.
\end{abstract}

\begin{IEEEkeywords}
Multi-source data fusion; 
Hyperspectral image; 
Synthetic aperture radar; 
Light detection and ranging;
Shared-pattern collaborative learning.
\end{IEEEkeywords}

\section{Introduction}

\IEEEPARstart{R}{emote} sensing image classification has become a fundamental task in Earth observation \cite{ahmad2022hsisurvey}. It has attracted considerable attention in recent years owing to its wide-ranging applications in environmental monitoring \cite{gcz26tgrs}, urban planning \cite{lv25ijde}, and disaster management \cite{sxy25ijdrr}. With the increasing availability of heterogeneous Earth observation data, the task has expanded from single-source pixel recognition toward multimodal scene understanding. However, different sensors measure distinct physical properties and exhibit different spatial, spectral, and noise characteristics, making reliable joint representation learning a persistent challenge.

\begin{figure}[!t]
\centering
\includegraphics[width=0.98\columnwidth]{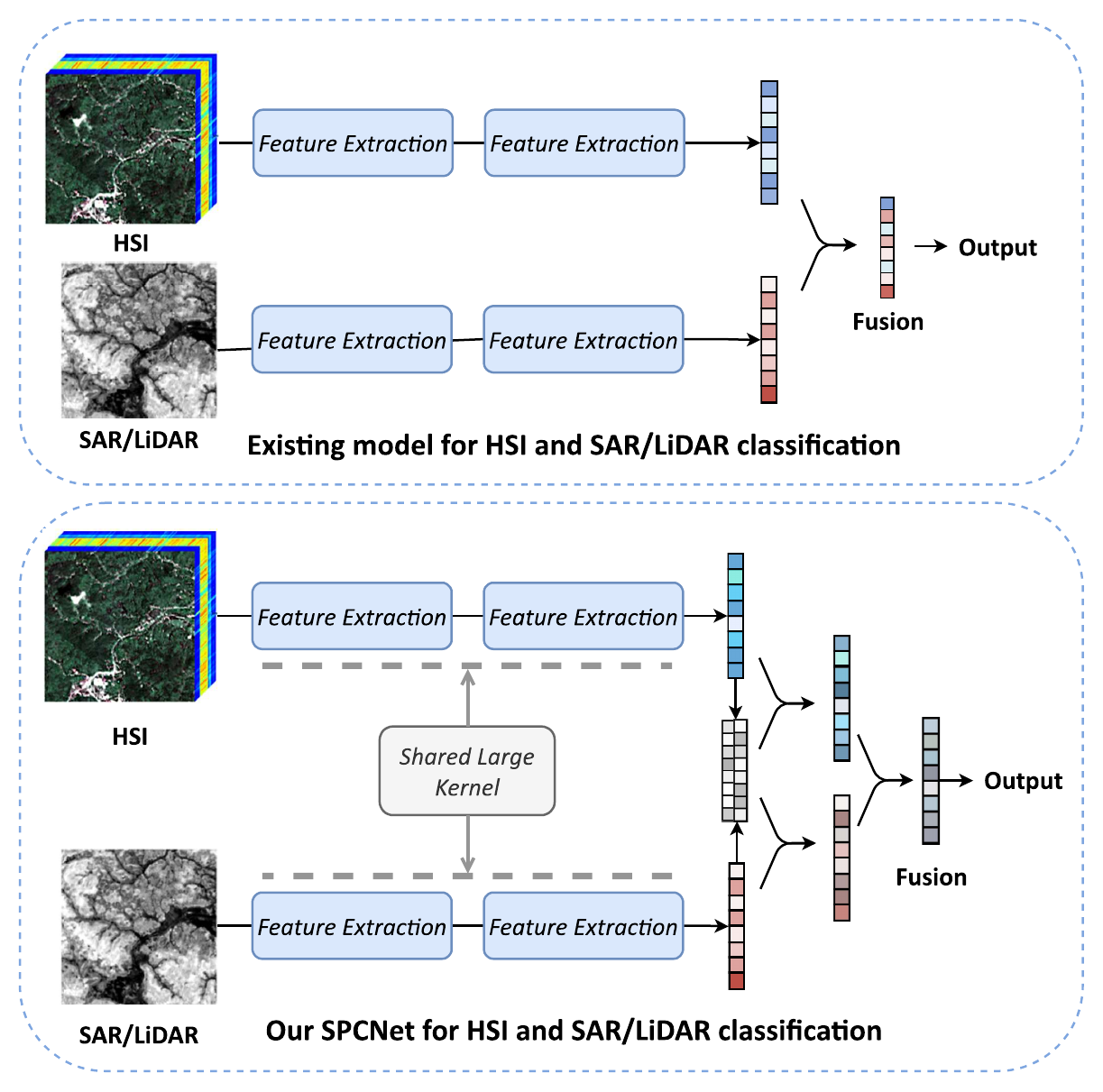}
\caption{Comparison between existing multi-source data classification models and the proposed SPCNet. SPCNet shares a large-kernel operator across the two modality branches to collaboratively learn common spatial patterns while retaining modality-specific features for adaptive fusion.}
\label{fig:motivation}
\end{figure}

Land-cover classification is not merely the recognition of isolated pixels. Natural and urban scenes contain objects at markedly different spatial scales, including compact buildings, elongated roads, fragmented vegetation, and mixed transition regions. Their spectra may exhibit high intra-class variability and strong inter-class similarity because of illumination changes, material mixtures, seasonal conditions, and sensor noise \cite{gao26pys}. An effective classifier must therefore jointly exploit diagnostic spectral signatures, local structures, and broader spatial context \cite{hao2021garnn,roy2021a2s2kresnet}.

Hyperspectral images (HSIs) acquire hundreds of contiguous spectral bands across the electromagnetic spectrum. Such rich spectral information provides strong capability for identifying complex objects \cite{hong26nrmp,li2019hsioverview,rasti2020feature,li25edip}, but neighboring bands are highly correlated and limited spatial resolution may blur object boundaries. The classification performance of HSI is therefore often constrained by spectral similarity among different land-cover classes, noise contamination, and spectral redundancy. Integrating HSI with other modalities has consequently become an effective strategy for improving classification performance \cite{li24m2unet}. Synthetic aperture radar (SAR) and Light Detection and Ranging (LiDAR) data are particularly attractive because of their distinct imaging mechanisms and rich structural information \cite{ghamisi2019fusionreview,xue2022selfsupervised,ljy23ssmae}. SAR captures geometric and textural characteristics under all-weather and all-day conditions, while LiDAR provides precise elevation and three-dimensional structural information. Their combination with HSI enables complementary spectral, scattering, and structural cues to be exploited jointly, as demonstrated by large multimodal benchmark studies \cite{hu2023mdas}. Motivated by these advantages, this paper investigates the joint classification of HSI and SAR/LiDAR data.

The benefit of multi-source observation does not simply arise from stacking additional channels. HSI records reflectance, SAR characterizes backscattering related to surface roughness and imaging geometry, whereas LiDAR directly measures elevation and three-dimensional structure. Their relationship is therefore established through land-cover semantics rather than homogeneous signal statistics. Moreover, source reliability is spatially nonuniform: HSI is affected by shadows and spectral mixing, SAR by speckle and layover, and LiDAR may be less informative over locally flat surfaces. Differences in spatial resolution, noise distribution, and physical response consequently create a substantial cross-modal gap \cite{houston2013,hong21,ding2022glt,hong2021diverse}. A successful fusion model should preserve modality-specific evidence while discovering structural regularities shared across sensors, rather than propagating redundant or unreliable responses through indiscriminate aggregation.

Multi-source classification methods have consequently evolved from decision-level fusion toward feature-level collaborative learning. Decision-level approaches combine predictions produced by independently trained classifiers and are relatively robust to heterogeneous feature dimensions, but their late interaction cannot fully exploit intermediate correlations between spectral, scattering, and elevation cues. Feature-level methods use parallel encoders and exchange information during representation learning, allowing complementary cues to participate directly in the construction of class-discriminative features \cite{xu2018tbcnn,hang2020coupledcnn}. CNN-based architectures provide strong local inductive biases, while attention mechanisms and Transformer-based models improve global interaction and long-range dependency modeling \cite{mohla2020fusatnet,yao2023exvit,zhang2023mtn}. Shared-and-specific representation learning and masked modeling further demonstrate that cross-source consistency and modality individuality should be considered jointly \cite{hong21,ljy23ssmae}. Nevertheless, increasing network depth or introducing global attention does not automatically guarantee that heterogeneous branches learn compatible spatial structures, nor does it determine which source should dominate at every location.

As illustrated in Fig.~\ref{fig:motivation}, existing multi-source models typically extract HSI and SAR/LiDAR features using independent operators before combining the resulting representations. Such separate processing can overlook the common spatial structures shared by co-registered observations. SPCB selectively shares a patch-spanning spatial operator across co-registered modalities and block depths, while modality-specific stems, identity channels, and adaptive local filtering retain source-dependent features. This design complements shared-specific and cross-modal learning \cite{hong21}.

Deep learning-based methods have become the mainstream solution for multi-source remote sensing classification. Existing approaches are primarily based on convolutional neural networks (CNNs) \cite{xu2018tbcnn,fang2022s2enet}, Transformers \cite{phz25jstars,lc25tgrs}, or their hybrid architectures to exploit complementary information. Despite their promising performance, two challenges remain insufficiently addressed: \textbf{1) \textit{Difficulty in simultaneously capturing long-range dependencies and shared spatial patterns between heterogeneous modalities.}} Many existing methods process different modalities independently or simply fuse them at a late stage, which limits their ability to discover common spatial structures across modalities in feature learning. Consequently, the shared contextual patterns and complementary structural cues across heterogeneous modalities remain insufficiently explored. \textbf{2) \textit{Limited region-adaptive feature fusion.}} Most methods adopt fixed or globally learned fusion schemes, which fail to dynamically adjust the contribution of each modality across different regions. Consequently, region-specific complementary information may be insufficiently exploited, while redundant or unreliable modality features can be introduced into the fused representation.

These two challenges are closely coupled. If the feature extractors of different sources model context using unrelated spatial operators, subsequent fusion must align not only heterogeneous physical measurements but also incompatible spatial responses. Conversely, even well-aligned representations can be degraded when a globally fixed fusion rule assigns excessive importance to an unreliable modality in a particular region. Co-registered HSI and SAR/LiDAR observations share object boundaries, neighborhood organization, and scene-level geometry, suggesting that common spatial patterns can serve as a stable inductive prior for cross-modal learning. At the same time, the appearance of these patterns remains content dependent, and the discriminative contribution of each source changes with land-cover type and local observation conditions. Therefore, the desired framework should combine shared contextual modeling with content-adaptive local representation, and should perform fusion according to spatially varying source reliability rather than a single scene-level preference.

\begin{figure*}
\centering
\includegraphics[width=6in]{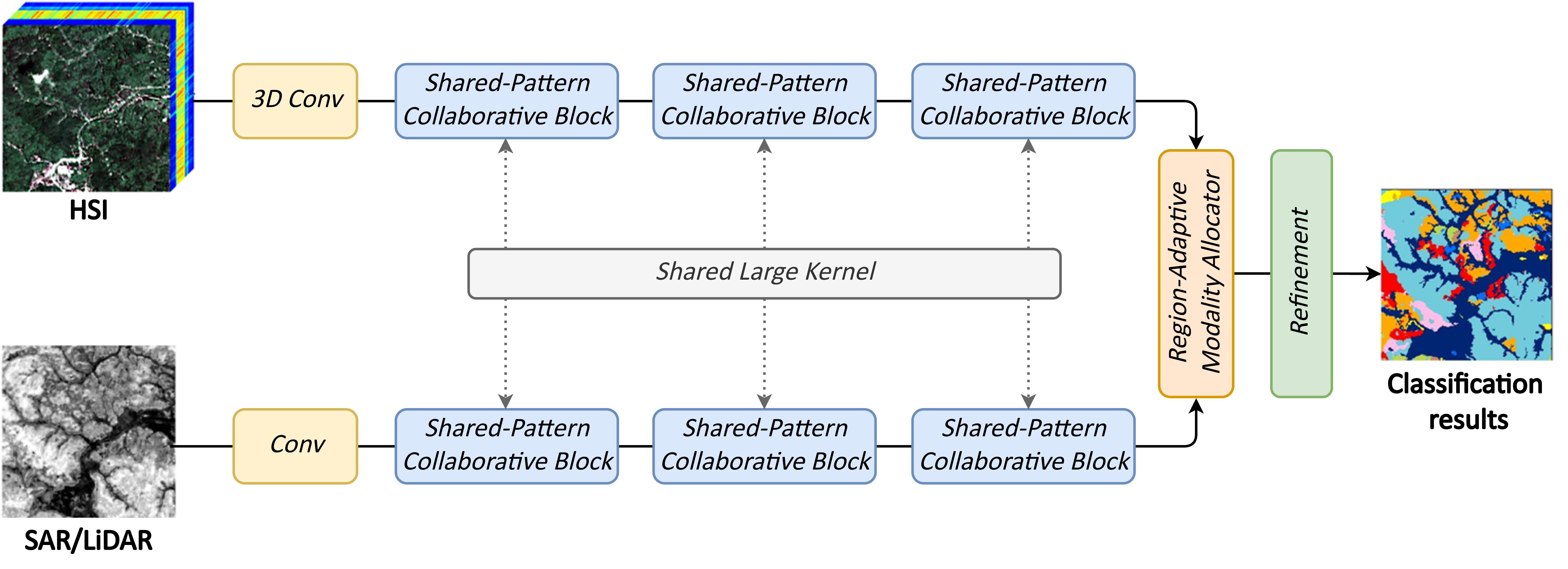}
\caption{Overview of the proposed \textbf{Shared-Pattern Collaborative Network (SPCNet)} for multi-source remote sensing data classification. The input multi-source data are processed by stacked Shared-Pattern Collaborative Blocks (SPCB) that exchange and learn common spatial-spectral representations through shared large kernels. Region-Adaptive Modality Allocator (RAMA) dynamically integrates complementary modality information to generate the classification map.}
\label{fig:frame}
\end{figure*}

To overcome these limitations, this paper proposes a \textbf{S}hared-\textbf{P}attern \textbf{C}ollaborative \textbf{Net}work (\textbf{SPCNet}). To promote effective contextual feature learning, we design \textit{Shared-Pattern Collaborative Block (SPCB)}, in which a shared large-kernel operator is applied to both branches to capture common spatial patterns and enhance long-range contextual modeling. Meanwhile, an input-adaptive dynamic convolution is introduced to strengthen content-aware local feature representation. Furthermore, we introduce \textit{Region-Adaptive Modality Allocator (RAMA)} to dynamically estimate the contribution of each modality in different spatial regions. Instead of using fixed fusion weights, RAMA adaptively assigns region-specific modality importance, allowing the network to emphasize the more reliable modality according to local scene characteristics.

The main contributions of this work are threefold:

\begin{itemize}

\item We design SPCB to enhance cross-modal collaborative contextual feature learning. By introducing a shared large-kernel operator, SPCB encourages different modalities to learn common spatial patterns, while the dynamic convolution branch provides content-adaptive feature modeling.

\item  We introduce RAMA for adaptive multi-source feature fusion. It dynamically assigns modality-specific weights to different spatial regions, allowing the network to emphasize the more reliable modality according to local scene characteristics.

\item Comprehensive experiments conducted on three benchmark datasets verify the effectiveness of the proposed SPCNet. In addition, we will publicly release the source code to facilitate the development of multi-source data fusion methods.

\end{itemize}

\section{Related Works}

\subsection{Multi-Source Remote Sensing Image Classification}

Multi-source remote sensing classification methods can be broadly grouped according to the stage at which heterogeneous information is integrated. Decision-level fusion trains separate classifiers for different sources and combines their outputs by voting, probability aggregation, or graph-based reasoning. This strategy avoids directly reconciling heterogeneous feature dimensions and can remain effective when one modality is missing or severely degraded. However, because the sources interact only after independent inference, the learned representations cannot use complementary information to resolve ambiguous samples during feature extraction. Feature-level fusion has therefore become the dominant paradigm. It usually employs modality-specific encoders followed by summation, concatenation, attention, or learned interaction modules, enabling spectral, spatial, scattering, and elevation cues to be optimized jointly \cite{xu2018tbcnn,hang2020coupledcnn,wu2022multimodalcnn}.

CNN-based methods established the basic dual-stream architecture for HSI and LiDAR/SAR joint classification. Coupled CNNs exchange information between modality branches, while spatial-spectral enhancement and recurrent attention models strengthen local feature interaction at multiple scales \cite{hang2020coupledcnn,fang2022s2enet,l2022a3clnn}. FusAtNet introduces dual attention to emphasize informative spectral-spatial responses \cite{mohla2020fusatnet}, and asymmetric or collaborative fusion networks model unequal source characteristics instead of treating the modalities identically \cite{li2022asyffnet,tang2024micf}. Depthwise feature interaction, fractional Gabor filtering, nearest-neighbor contrastive learning, and cross-scale attention further improve multiscale and cross-source discrimination \cite{gao2022dfinet,zhao2022fgcn,wang2023nncl,gao2023csma}. These methods provide effective local representation and relatively stable optimization. Nevertheless, standard convolutions operate within limited receptive fields. Enlarging the network depth can propagate information over wider regions, but indirect long-range modeling may weaken boundary details and increase redundant computation.

Transformer and hybrid architectures address this limitation by modeling global token relationships. Multimodal Transformer networks and ExViT establish long-range dependencies through self-attention and cross-modal attention \cite{zhang2023mtn,yao2023exvit,roy2023mft}, whereas hierarchical CNN--Transformer models combine convolutional locality with global interaction \cite{zhao2022hct,li2023macn}. Deep symmetric fusion, local information interaction, multiscale mixed CNN--Transformer modeling, and head selection further tailor global representation learning to multimodal remote sensing data \cite{chang2024dsft,zhang2023liit,sun2024mat,ni2024mhst}. Quaternion, high-order, and progressive cross-fusion mechanisms further enrich the interaction between heterogeneous representations \cite{hu2025gccqtnet,ni2025chnet,phz25jstars}. These architectures are effective for modeling nonlocal context, but dense attention may introduce considerable computation and can establish correlations between semantically unrelated positions. More importantly, global interaction alone does not ensure that different branches respond to the same object structures in a consistent manner.

Recent studies have therefore investigated shared-and-specific representations, masked pretraining, mutual guidance, and iterative cross-source refinement. S2FL learns shared and modality-specific subspace projections \cite{hong21}. SS-MAE uses spatial-spectral masked modeling to learn multi-source representations \cite{ljy23ssmae}.  Mutual-guidance or spectral-correlation networks progressively refine one source with information from another \cite{zhang2025mgmnet,gcz26tgrs}. Style separation, frequency-domain fusion, metric learning, and multiscale state-space modeling address cross-source discrepancies from complementary perspectives \cite{han2023ssmunet,ni2024fdbn,song2023hdml,gao2025msfmamba}. These approaches demonstrate the importance of cross-source consistency.

\subsection{Context Modeling and Adaptive Multi-Source Fusion}

Context modeling determines whether a classifier can distinguish objects using both local appearance and surrounding structure. Conventional small-kernel CNNs are effective at preserving local details, but their effective receptive fields grow slowly with network depth. Large-kernel CNNs broaden spatial support, while selective large-kernel, dynamic, and cross-guided filters adapt contextual responses to input content or another modality \cite{ding2022replknet,li2025lsknet,chen2020dynamicconv,yang2019condconv,zhang2025mgmnet}. Cross-modal interaction and selected parameter sharing also promote source consistency \cite{hang2020coupledcnn,tang2024micf}. These directions leave a joint challenge: integrating long-range dependencies with a spatial prior shared across registered modalities and network depths, while retaining input-sensitive local filtering. SPCB addresses this by embedding the shared prior in feature extraction and pairing it with filtering that adapts to each input. The shared prior captures recurring scene geometry across modalities and network depths; adaptive filtering models local patterns that vary by source and input.

The need for shared contextual modeling follows from the spatial correspondence of registered observations. Although HSI, SAR, and LiDAR describe different physical attributes, roads, buildings, vegetation parcels, and object boundaries occupy approximately the same spatial locations. Their low-level intensity patterns differ, but their neighborhood organization and larger geometric structures are semantically related.

Fusion strategy is equally important because complementary information is not uniformly reliable. Addition and concatenation assume fixed source roles and rely on subsequent convolutions to suppress irrelevant responses. Channel attention and cross-attention learn data-dependent interactions \cite{mohla2020fusatnet,tang2024micf,phz25jstars}, while intensity constraints, mutual guidance, and global weighting further improve source selection \cite{lc25tgrs,zhang2025mgmnet}. GMU gates modality contributions at the representation level \cite{arevalo2017gmu}. Recent remote-sensing methods further use cross-modal attention and semantic guidance, mutual guidance, or graph-based interaction with feature-dependent fusion \cite{tang2024micf,zhang2025mgmnet,gao2024ien}. These mechanisms support adaptive interaction, but spatial saliency within one representation does not itself compare the reliability of competing sources, and feature-wise modality gates do not explicitly allocate source contributions by location. Because shared scene geometry does not make modality evidence equally discriminative in every region, fusion must retain this spatial variation. RAMA addresses this requirement by deriving spatially varying modality weights from joint HSI-auxiliary features after shared-pattern extraction, so that source contributions can adapt to local content. This regional variation is reflected in the complementary cues provided by each source: LiDAR can distinguish spectrally similar objects with different elevations, HSI can discriminate materials in regions with similar geometry, and SAR can provide structural cues when optical observations are unreliable.

Accordingly, robust multi-source fusion requires two complementary capabilities: cross-source feature extractors should learn compatible contextual patterns, and the fusion module should assign source contributions according to regional content.

\section{Methodology}

The overall framework of the proposed Shared-Pattern Collaborative Network (SPCNet) is illustrated in Fig. \ref{fig:frame}. SPCNet contains two parallel modality-specific branches: the HSI branch uses PCA followed by a 3D convolutional stem, whereas the SAR/LiDAR branch uses a 2D convolutional stem. Their features are subsequently refined by stacked SPCBs and integrated by RAMA.

Let $\mathbf{X}_h\in\mathbb{R}^{H\times W\times B}$ and $\mathbf{X}_x\in\mathbb{R}^{H\times W\times C_x}$ denote the spatially aligned HSI and auxiliary-source patches, respectively, where $B$ is the number of original hyperspectral bands and $C_x$ is the number of SAR/LiDAR channels. PCA retains the leading $N_{\rm PC}$ components of the HSI before 3D convolution extracts joint spectral-spatial features, whereas the 2D auxiliary branch extracts structural, scattering, or elevation features. The two stems retain modality-specific low-level statistics while projecting both sources into compatible feature spaces.

\begin{figure}
\centering
\includegraphics[width=\columnwidth]{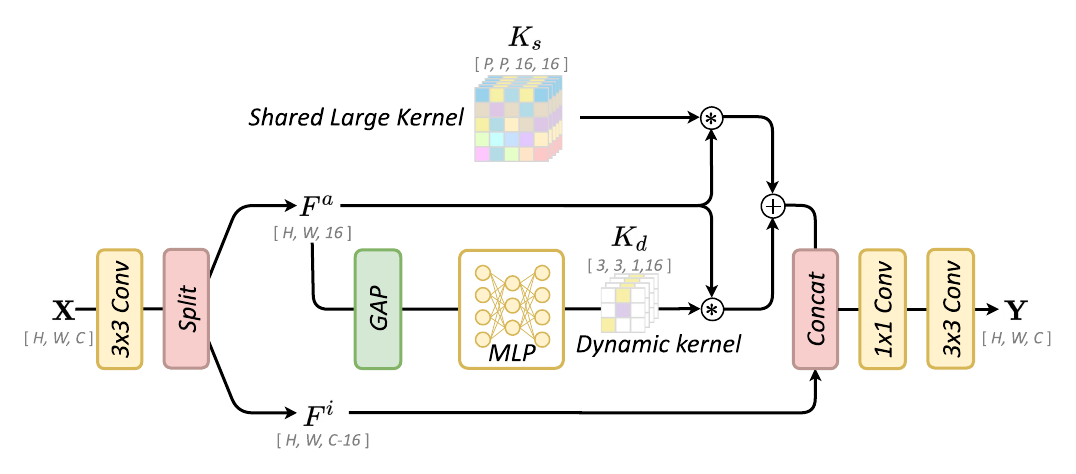}
\caption{Details of the Shared-Pattern Collaborative Block (SPCB).}
\label{fig:spcb}
\end{figure}

The encoded features are successively processed by stacked Shared-Pattern Collaborative Blocks (SPCBs) and the Region-Adaptive Modality Allocator (RAMA). The internal structure of SPCB is shown in Fig.~\ref{fig:spcb}. SPCB learns common spatial patterns while preserving source-dependent features, and RAMA produces the fused representation $\mathbf{F}_{fus}$ for pixel-wise classification. The two modules and their end-to-end optimization are described below.

\subsection{Shared-Pattern Collaborative Block (SPCB)}

Given an input feature $\mathbf{X}\in \mathbb{R}^{H\times W\times C}$, SPCB first applies a $3\times3$ convolution to obtain feature $\mathbf{F}$. Then, $\mathbf{F}$ is split along the channel dimension into $\mathbf{F}^{a}\in \mathbb{R}^{H\times W\times 16}$ and $\mathbf{F}^{i}\in \mathbb{R}^{H\times W\times (C-16)}$. Here, $\mathbf{F}^a$ is used as the active branch for shared-pattern learning and dynamic kernel generation, while $\mathbf{F}^i$ is preserved as an identity branch and later concatenated with the enhanced feature. Because these channels bypass the shared and dynamic filters before rejoining the output projection, they carry shallow responses into later mixing. Cross-depth reuse therefore acts on progressively refined features that retain earlier cues, while the dynamic branch adapts local filtering to each input.

Restricting the active branch to 16 channels keeps the large-kernel and dynamic operations lightweight. The remaining $C-16$ channels form a direct information path, reducing the risk that repeated adaptive filtering removes fine source-specific details. The channel split thus provides a practical balance between contextual enhancement and feature preservation.

\textbf{Shared Large-Kernel Pattern.} The upper branch uses a shared large convolution kernel $K_s \in \mathbb{R}^{P\times P\times 16\times 16}$ to model broad spatial context, where $P$ denotes the side length of the input image patch. Since the same kernel $K_s$ is applied to both HSI and SAR/LiDAR features, the two branches are encouraged to extract spatial structures under a unified convolutional pattern. Through kernel sharing, SPCB establishes an implicit interaction mechanism between the two modalities, allowing the network to align and compare their spatial responses in a common feature space. The shared-pattern response is computed as:

\begin{equation}
  \mathbf{F}^a_s = K_s * \mathbf{F}^a,
\end{equation}
where $*$ denotes convolution, and $\mathbf{F}^{a}_{s}\in \mathbb{R}^{H\times W\times 16}$. The large kernel helps the network capture long-range context and shared spatial structures. With stride one, the operator acts on the patch grid. Reflection padding of $(P-1)/2$ pixels per side preserves the output size, while the support centered on the target pixel spans the unpadded input patch. The kernel side length equals the patch side length $P$.

Compared with repeatedly stacking small convolutions to approximate a large receptive field, the $P\times P$ operator directly aggregates the spatial context of the input patch within one collaborative unit. Sharing its parameters constrains the two source branches to analyze that neighborhood with the same spatial template. Reusing the operator at different depths further maintains a consistent structural prior as the features evolve from low-level responses to more semantic representations.

The shared operator does not require the two modalities to produce identical feature values. Instead, it constrains the spatial relation used to process them, while the identity and dynamic branches retain modality-dependent evidence and local appearance variation. This selective sharing is important for aligned multi-source observations, whose geometry is common but whose radiometric responses remain heterogeneous.

\textbf{Dynamic Kernel Generation.} SPCB introduces a dynamic kernel branch to generate an adaptive convolution kernel according to the input feature. First, global average pooling is applied to $\mathbf{F}^a$, and MLP is used to predict the dynamic kernel as:
\begin{equation}
  K_d= \mathrm{MLP}(\mathrm{GAP}(\mathbf{F}^a)),
\end{equation}
where the generated lightweight dynamic kernel is $K_d\in\mathbb{R}^{3\times3\times1\times16}$. For each channel of $\mathbf{F}^a$, a corresponding $3\times3$ dynamic filter is used as:
\begin{equation}
\mathbf{F}^{a}_{d} = \mathrm{DyDConv}(\mathbf{F}^{a}; K_d),  
\end{equation}
where $\mathrm{DyDConv}$ denotes dynamic depth-wise convolution. The dynamic depth-wise convolution is more efficient while still providing input-adaptive feature modeling capability. 

The global descriptor produced by GAP summarizes the current sample and allows the MLP to generate a different group of channel-wise filters for different inputs. Because each $3\times3$ filter operates depth-wise, the dynamic branch avoids the parameter and computation cost of a dense input-dependent convolution. It therefore complements the fixed shared large kernel with efficient local adaptation rather than duplicating its long-range modeling role.

\textbf{Collaborative Fusion.} The outputs of the shared large-kernel branch and the dynamic kernel branch are then combined by element-wise addition, and then concatenated with $\mathbf{F}^i$ as:
\begin{equation}
  \mathbf{F}^c= \mathrm{Concat}(\mathbf{F}^a_s+\mathbf{F}^a_d, ~ \mathbf{F}^i).
\end{equation}
Element-wise addition first combines the common-context response and adaptive local response in the same 16-channel space. Concatenation with $\mathbf{F}^i$ then restores the preserved identity information. Finally, a $1\times1$ convolution mixes the enhanced and identity channels, and a $3\times3$ convolution refines their local spatial consistency to generate the output feature:
\begin{equation}
\mathbf{Y}=\mathrm{Conv}_{3\times3}\bigl(\mathrm{Conv}_{1\times1}(\mathbf{F}^c)\bigr).
\end{equation}
This two-step projection enables information exchange across channels before local spatial refinement and prepares a compatible feature representation for the next SPCB or the subsequent fusion stage.

\subsection{Region-Adaptive Modality Allocator (RAMA)}

\begin{figure}
\centering
\includegraphics[width=\columnwidth]{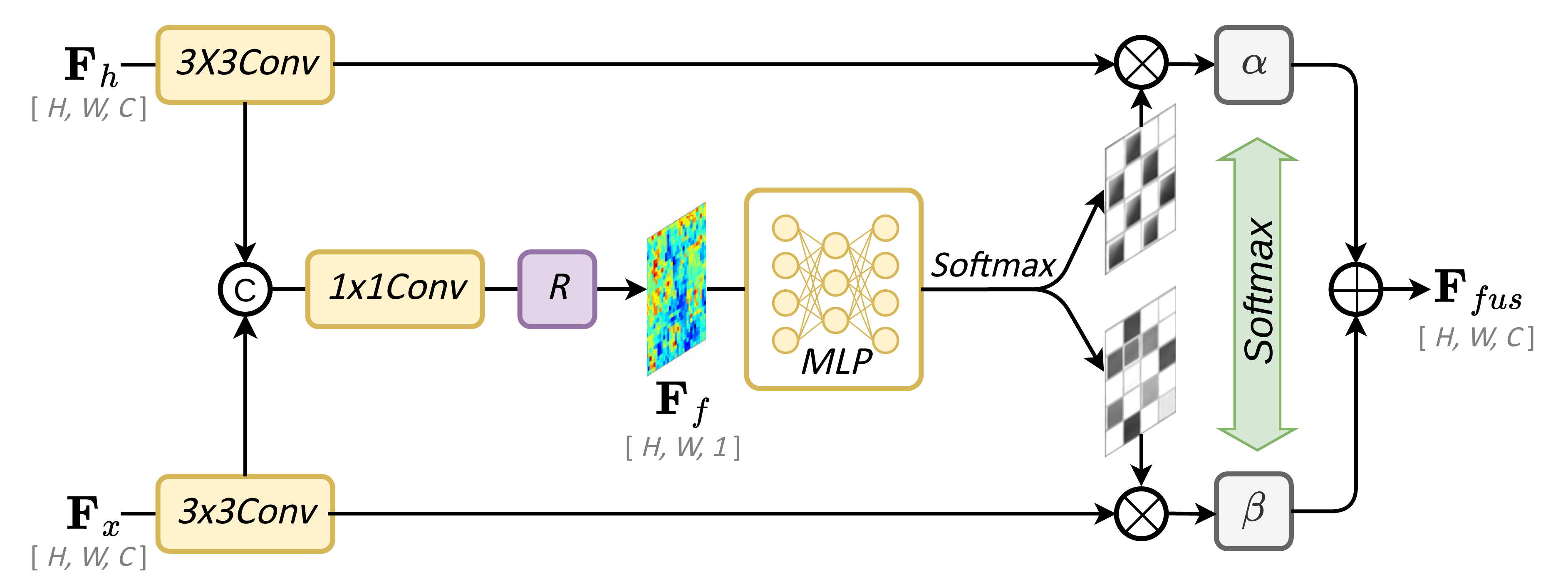}
\caption{Details of the Region-Adaptive Modality Allocator (RAMA).}
\label{fig:rama}
\end{figure}

The Region-Adaptive Modality Allocator (RAMA), illustrated in Fig.~\ref{fig:rama}, is designed to adaptively allocate the contributions of multi-source features. RAMA learns spatially varying modality weights and performs region-aware fusion. Instead of assigning a fixed global importance to each modality, RAMA dynamically determines which modality should contribute more at each spatial location.

For a mini-batch of $M$ co-registered patch pairs, the input features $\mathbf{F}_h,\mathbf{F}_x\in\mathbb{R}^{M\times C\times H\times W}$ are independently processed by $3\times3$ convolutions to obtain $\hat{\mathbf{F}}_h$ and $\hat{\mathbf{F}}_x$. Their concatenation has shape $M\times2C\times H\times W$, and a $1\times1$ projection produces $\mathbf{F}_r$ of the same shape.
\begin{equation}
  \mathbf{F}_{cat} = \mathrm{Concat}(\hat{\mathbf{F}}_h, \hat{\mathbf{F}}_x),
\end{equation}
\begin{equation}
  \mathbf{F}_r= \mathrm{Reshape}(\mathrm{Conv_{1\times1}}(\mathbf{F}_{cat})). 
\end{equation}

Flattening the spatial axes gives $M\times2C\times N_s$, where $N_s=HW$. An MLP produces two logits per feature pixel, with grid shape $M\times2\times N_s$. Softmax over modalities and reshaping yield pixelwise maps $[\mathbf{A}_h,\mathbf{A}_x]\in\mathbb{R}^{M\times2\times H\times W}$, each of shape $M\times1\times H\times W$:
\begin{equation}
\begin{aligned}
\relax[\mathbf{A}_h(i,j),\mathbf{A}_x(i,j)]
 &=\mathrm{Softmax}_{m}\bigl(\mathrm{MLP}(\mathbf{F}_r(i,j))\bigr),\\
\mathbf{A}_h(i,j)+\mathbf{A}_x(i,j)&=1.
\end{aligned}
\end{equation}

Each map is broadcast over $C$ channels to weight its refined feature. The maps allocate contributions at each patch feature pixel, while input-independent global coefficients $\alpha$ and $\beta$ calibrate their overall contributions.
\begin{equation}
\hat{\mathbf{F}}'_h = \mathbf{A}_h \odot \hat{\mathbf{F}}_h,
~~~ \hat{\mathbf{F}}'_x = \mathbf{A}_x \odot \hat{\mathbf{F}}_x,
\end{equation}

\begin{equation}
\mathbf{F}_{fus}=\alpha\hat{\mathbf{F}}'_h+\beta\hat{\mathbf{F}}'_x,\qquad \alpha+\beta=1.
\end{equation}

\subsection{Optimization Objective}

For an input sample with $K$ land-cover categories, the classification head converts $\mathbf{F}_{fus}$ into the predicted class-probability vector $\mathbf{p}_n$. SPCNet is trained end-to-end using the standard cross-entropy objective:
\begin{equation}
\mathcal{L}_{\mathrm{ce}}=-\frac{1}{N}\sum_{n=1}^{N}\sum_{k=1}^{K} y_{n,k}\log p_{n,k},
\end{equation}
where $N$ is the number of labeled training samples in a mini-batch and $y_{n,k}$ is the one-hot ground-truth indicator. The classification loss jointly supervises the modality-specific encoders, SPCBs, dynamic kernel generator, RAMA, and final prediction head, so the shared patterns and regional allocations are learned specifically for discriminative land-cover recognition. During back-propagation, the shared spatial operator, dynamic local filters, and allocation maps are optimized jointly rather than through separate auxiliary objectives, ensuring that each mechanism is directly guided by the final classification error.

\section{Experimental Results and Analysis}

\subsection{Experimental Setup and Dataset Description}

To evaluate the effectiveness and generality of SPCNet, experiments were conducted on three benchmark multi-source remote sensing datasets: Houston2013, Augsburg, and Houston2018. All methods were implemented in PyTorch and trained on an NVIDIA H100 GPU with 80 GB of memory. The same training and testing partitions were used for all compared methods to ensure a fair evaluation. The Houston datasets pair HSI with LiDAR-derived structural information, whereas Augsburg combines HSI with polarimetric SAR (PolSAR) observations. These complementary settings allow the proposed network to be evaluated under both spectral--elevation and spectral--scattering fusion conditions.

\begin{table}[t]
\centering
\caption{Number of training and test samples for the Houston2013 dataset.}
\label{tab:h13split}
\footnotesize
\renewcommand{\arraystretch}{0.80}
\setlength{\tabcolsep}{2.8pt}
\begin{tabular*}{0.84\columnwidth}{@{\extracolsep{\fill}}r l c r r}
\toprule
No. & Name & Color & Train & Test \\
\midrule
1 & Healthy grass & \classswatch{000080} & 198 & 1053 \\
2 & Stressed grass & \classswatch{0000C8} & 190 & 1064 \\
3 & Synthetic grass & \classswatch{001EFF} & 192 & 505 \\
4 & Trees & \classswatch{0068FF} & 188 & 1056 \\
5 & Soil & \classswatch{00A5F5} & 186 & 1056 \\
6 & Water & \classswatch{00DDEB} & 182 & 143 \\
7 & Residential & \classswatch{36EDC2} & 196 & 1072 \\
8 & Commercial & \classswatch{80F06B} & 191 & 1053 \\
9 & Road & \classswatch{C5F52A} & 193 & 1059 \\
10 & Highway & \classswatch{FFF200} & 191 & 1036 \\
11 & Railway & \classswatch{FFA000} & 181 & 1054 \\
12 & Parking lot 1 & \classswatch{FF6A00} & 192 & 1041 \\
13 & Parking lot 2 & \classswatch{FF1D00} & 184 & 285 \\
14 & Tennis court & \classswatch{D40000} & 181 & 247 \\
15 & Running track & \classswatch{990000} & 187 & 473 \\
\midrule
\multicolumn{3}{c}{Total} & 2832 & 12197 \\
\bottomrule
\end{tabular*}
\end{table}

\begin{table}[t]
\centering
\caption{Number of training and test samples for the Augsburg dataset.}
\label{tab:augsplit}
\footnotesize
\renewcommand{\arraystretch}{0.80}
\setlength{\tabcolsep}{2.8pt}
\begin{tabular*}{0.80\columnwidth}{@{\extracolsep{\fill}}r l c r r}
\toprule
No. & Name & Color & Train & Test \\
\midrule
1 & Forest & \classswatch{18A527} & 146 & 13361 \\
2 & Residential area & \classswatch{D9D9D9} & 264 & 30065 \\
3 & Industrial area & \classswatch{D65359} & 21 & 3830 \\
4 & Low plants & \classswatch{00C83A} & 248 & 26609 \\
5 & Allotment & \classswatch{D7E300} & 52 & 523 \\
6 & Commercial area & \classswatch{BB58D5} & 7 & 1638 \\
7 & Water & \classswatch{083DFA} & 23 & 1507 \\
\midrule
\multicolumn{3}{c}{Total} & 761 & 77533 \\
\bottomrule
\end{tabular*}
\end{table}

\begin{table}[t]
\centering
\caption{Number of training and test samples for the Houston2018 dataset.}
\label{tab:h18split}
\footnotesize
\renewcommand{\arraystretch}{0.80}
\setlength{\tabcolsep}{2.4pt}
\begin{tabular*}{0.90\columnwidth}{@{\extracolsep{\fill}}r l c r r}
\toprule
No. & Name & Color & Train & Test \\
\midrule
1 & Healthy grass & \classswatch{24C83A} & 1000 & 38196 \\
2 & Stressed grass & \classswatch{91F32A} & 1000 & 129008 \\
3 & Artificial turf & \classswatch{008C82} & 1000 & 1736 \\
4 & Evergreen trees & \classswatch{17872B} & 1000 & 53322 \\
5 & Deciduous trees & \classswatch{24595B} & 1000 & 19172 \\
6 & Bare earth & \classswatch{9B4F12} & 1000 & 17064 \\
7 & Water & \classswatch{0DD9E6} & 500 & 564 \\
8 & Residential buildings & \classswatch{FFFFFF} & 1000 & 157995 \\
9 & Non-residential buildings & \classswatch{D0D0D0} & 1000 & 893769 \\
10 & Roads & \classswatch{F91011} & 1000 & 182283 \\
11 & Sidewalks & \classswatch{A0A0A0} & 1000 & 135035 \\
12 & Crosswalks & \classswatch{626262} & 1000 & 5059 \\
13 & Major thoroughfares & \classswatch{8B0000} & 1000 & 184438 \\
14 & Highways & \classswatch{D26A00} & 1000 & 38438 \\
15 & Railways & \classswatch{FFA000} & 1000 & 26748 \\
16 & Paved parking lots & \classswatch{FFFF00} & 1000 & 44932 \\
17 & Unpaved parking lots & \classswatch{D7B90C} & 250 & 337 \\
18 & Cars & \classswatch{E500E5} & 1000 & 25289 \\
19 & Trains & \classswatch{0000F5} & 1000 & 20479 \\
20 & Stadium seats & \classswatch{3FCBC4} & 1000 & 26296 \\
\midrule
\multicolumn{3}{c}{Total} & 18750 & 2000160 \\
\bottomrule
\end{tabular*}
\end{table}

\textbf{Houston2013 Dataset.} The Houston2013 dataset \cite{houston2013} was released for the 2013 IEEE GRSS Data Fusion Contest and covers the University of Houston campus and its surrounding urban area. The airborne HSI has a spatial size of $349 \times 1905$ pixels and 144 spectral bands spanning 380--1050 nm, with a ground sample distance of 2.5 m. The co-registered LiDAR data provide complementary elevation cues for buildings, vegetation, and transportation infrastructure. The scene contains 15 land-cover classes, including several visually similar paved and built-up categories as well as narrow road and railway structures. It therefore provides a representative urban setting for assessing whether cross-modal contextual modeling can preserve object continuity while separating classes with similar spectral responses.

\textbf{Augsburg Dataset.} The Augsburg dataset \cite{hong21} covers the vicinity of Augsburg, Germany, and contains HySpex HSI together with Sentinel-1 PolSAR data. Both modalities are provided at a 30-m ground sample distance with a spatial size of $332 \times 485$ pixels. The HSI contains 180 bands from 0.4 to 2.5 $\mu$m, while the SAR modality includes VV and VH intensity together with the real and imaginary components of the off-diagonal PolSAR covariance term. Its ground-truth labels are derived from OpenStreetMap and describe seven urban and peri-urban classes. The coarse spatial resolution, heterogeneous land use, and highly uneven class distribution make this dataset particularly useful for testing whether a fusion model can exploit scattering information without amplifying source-specific noise or minority-class ambiguity.

\textbf{Houston2018 Dataset.} The Houston2018 dataset \cite{hong21} was introduced in the 2018 IEEE GRSS Data Fusion Contest and covers the University of Houston campus together with neighboring urban areas. The original release provides HSI, LiDAR, and optical observations; following the multi-source HSI--LiDAR setting, only the HSI and LiDAR modalities are used in this work. The HSI contains 48 bands over the 380--1050 nm range at 1-m spatial resolution, while the LiDAR data have a finer native spatial resolution and are spatially aligned with the HSI for classification. This larger and more diverse scene includes 20 land-cover categories spanning vegetation, buildings, roads, parking areas, railways, and small objects such as cars and trains. Its extensive and strongly imbalanced labeled set provides a demanding test of generalization across complex urban spatial layouts.

\begin{figure*}
\centering
\includegraphics[width=0.64\textwidth]{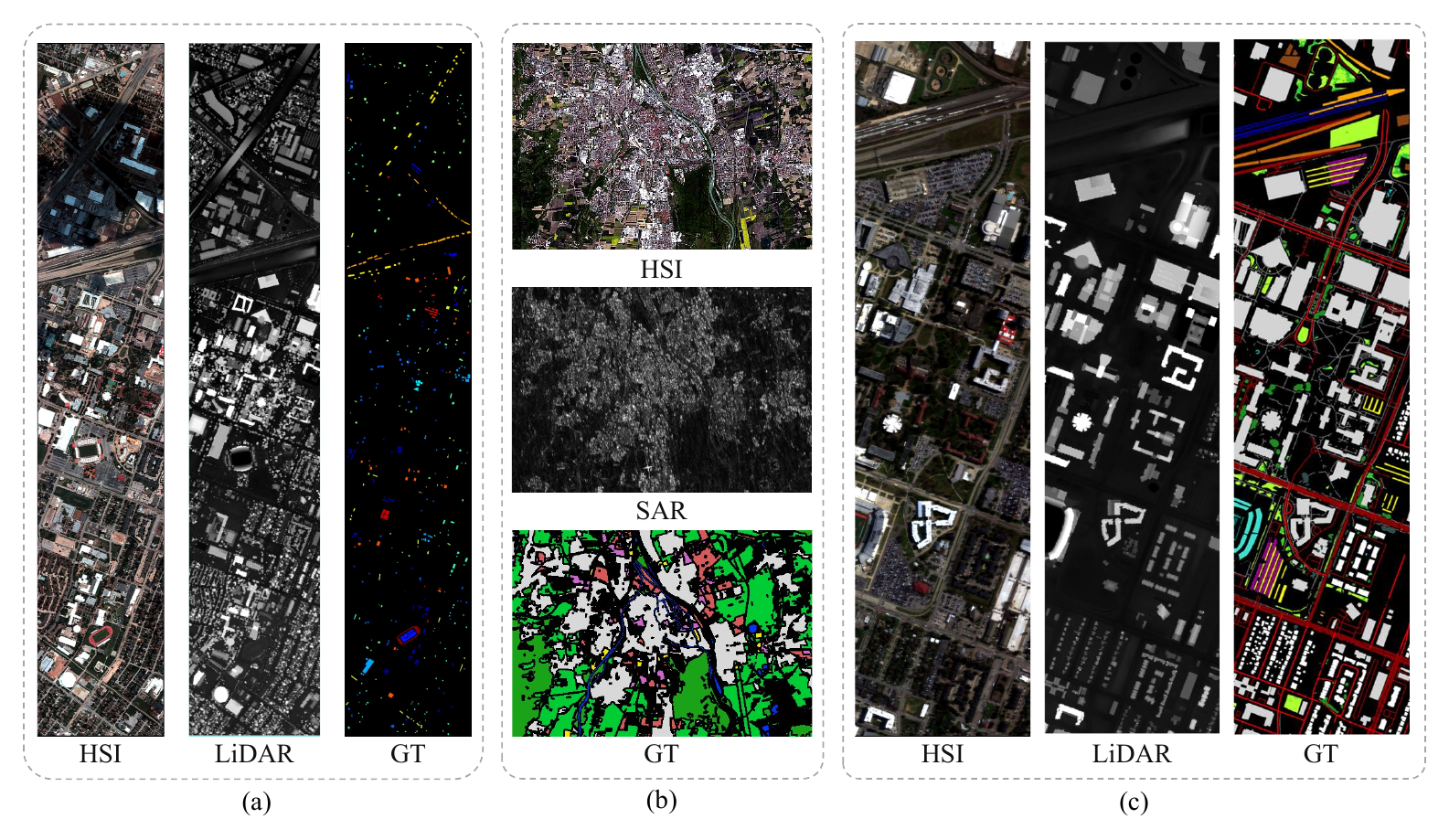}
\caption{Visualization of the three multi-source datasets. (a) Houston2013 dataset. (b) Augsburg dataset. (c) Houston2018 dataset.}
\label{fig:datasets}
\end{figure*}

The class-wise training and testing samples of Houston2013, Augsburg, and Houston2018 are listed in Tables~\ref{tab:h13split}, \ref{tab:augsplit}, and \ref{tab:h18split}, respectively. Fig.~\ref{fig:datasets} presents the false-color HSI, the corresponding SAR/LiDAR data, and the ground-truth maps of the three datasets.

After normalization, HSI is projected by PCA and each labeled target pixel defines a co-registered centered patch pair supervised by its center label. Reflection padding of $(P-1)/2$ pixels per side preserves output size. Both modalities receive the same horizontal or vertical flip with probability 0.5. All methods use these input settings and the same train/test partitions and optimization settings (Adam, initial learning rate $2\times10^{-4}$, weight decay 0, batch size 128, 35 epochs, dropout 0.05, and seed 6). We compare our SPCNet with ten state-of-the-art methods, including FusAtNet \cite{mohla2020fusatnet}, S2ENet \cite{fang2022s2enet}, AsyFFNet \cite{li2022asyffnet}, ExViT \cite{yao2023exvit}, HCT \cite{zhao2022hct}, MACN \cite{li2023macn}, MICF-Net \cite{tang2024micf}, GCCQTNet \cite{hu2025gccqtnet}, CHNet \cite{ni2025chnet}, and MGMNet \cite{zhang2025mgmnet}. To quantitatively assess the classification performance, the overall accuracy (OA), average accuracy (AA), and Kappa coefficient are employed for evaluation.

\begin{figure}[h]
\centering
\includegraphics[width=0.62\columnwidth]{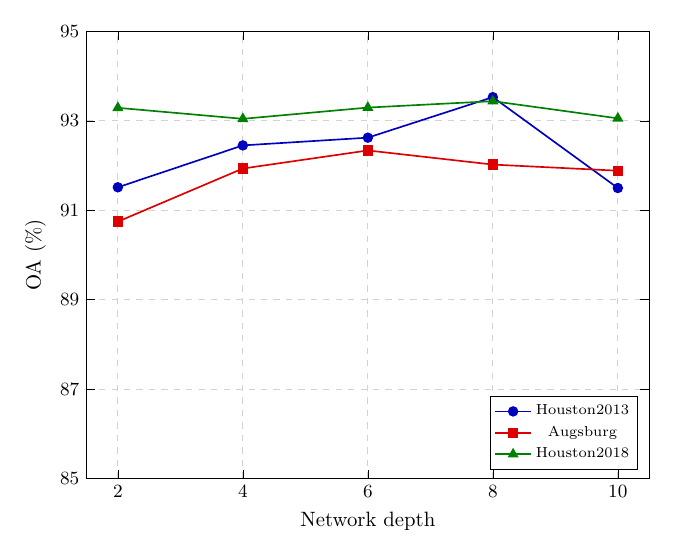}
\caption{Depth of SPCB Layers.}
\label{fig:param-depth}
\end{figure}

\begin{figure}[h]
\centering
\includegraphics[width=0.62\columnwidth]{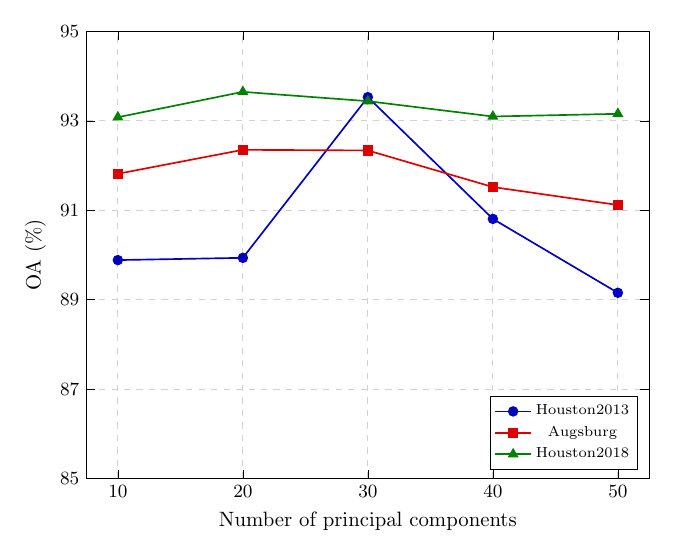}
\caption{Number of Principal Components for HSI.}
\label{fig:param-pca}
\end{figure}

\begin{figure}[h]
\centering
\includegraphics[width=0.62\columnwidth]{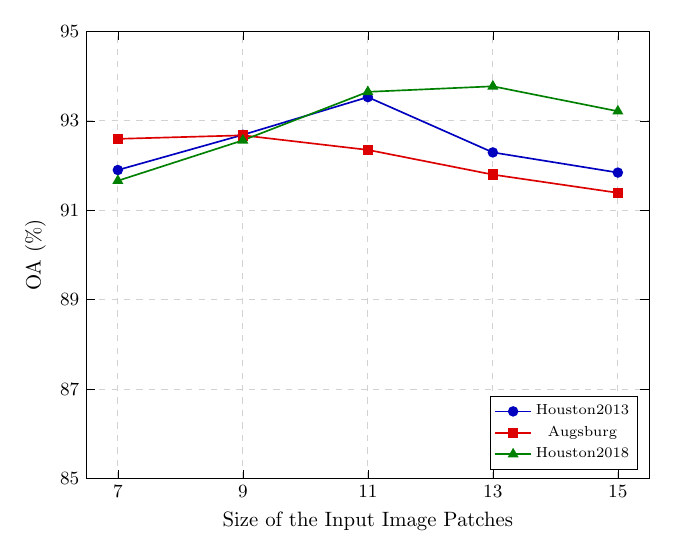}
\caption{Size of the Input Image Patches.}
\label{fig:param-window}
\end{figure}

\subsection{Analysis of Important Parameters}

\textbf{Depth of SPCB Layers.} Fig. \ref{fig:param-depth} shows the OA variation as the number of SPCB layers increases from 2 to 10. The best results are obtained with eight layers on Houston2013 (93.53\%) and Houston2018 (93.44\%), whereas Augsburg reaches its highest OA of 92.34\% with six layers. Increasing the depth initially improves the representation of shared spatial patterns and cross-source interactions. However, further stacking does not provide a consistent benefit, because redundant feature interactions can increase optimization difficulty and computation without contributing discriminative context. The selected depths are 8, 6, and 8 for Houston2013, Augsburg, and Houston2018, respectively.

\textbf{Number of Principal Components for HSI.} Fig. \ref{fig:param-pca} evaluates the number of retained principal components. On Houston2013, OA increases sharply to 93.53\% when 30 components are retained, while 40 and 50 components introduce redundant spectral information and reduce performance. Augsburg and Houston2018 achieve their best OAs of 92.35\% and 93.65\%, respectively, with 20 components. The different optima reflect the distinct spectral redundancy and scene composition of the three datasets. Accordingly, 30 principal components are retained for Houston2013, whereas 20 components are retained for Augsburg and Houston2018.

\textbf{Size of the Input Image Patches.} Fig. \ref{fig:param-window} studies the input patch size. Houston2013 reaches 93.53\% OA with an 11-pixel patch, Augsburg reaches 92.68\% with a 9-pixel patch, and Houston2018 reaches 93.77\% with a 13-pixel patch. Small patches provide insufficient contextual evidence for neighboring land-cover categories, whereas excessively large patches can include unrelated pixels and weaken boundary discrimination. The different optima therefore reflect differences in object scale and spatial heterogeneity among the three scenes.

\subsection{Performance Comparison}

\begin{figure*}[t]
\centering
\includegraphics[width=0.98\textwidth]{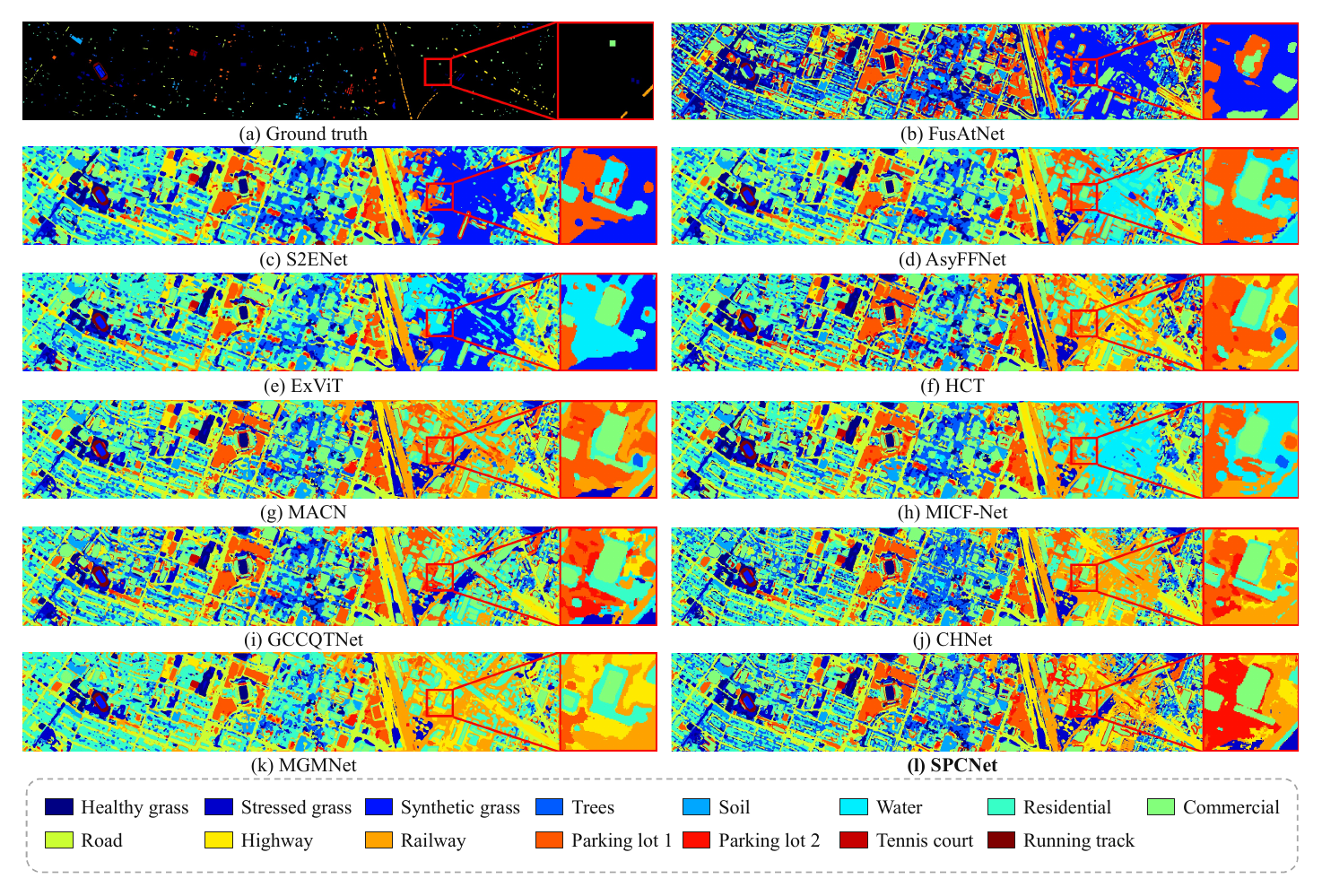}
\caption{Visualized classification results of different methods for the Houston2013 dataset. (a) Ground truth. (b) FusAtNet. (c) S2ENet. (d) AsyFFNet. (e) ExViT. (f) HCT. (g) MACN. (h) MICF-Net. (i) GCCQTNet. (j) CHNet. (k) MGMNet. (l) SPCNet.}
\label{fig:h13vis}
\end{figure*}

\begin{figure*}
\centering
\includegraphics[width=0.98\textwidth]{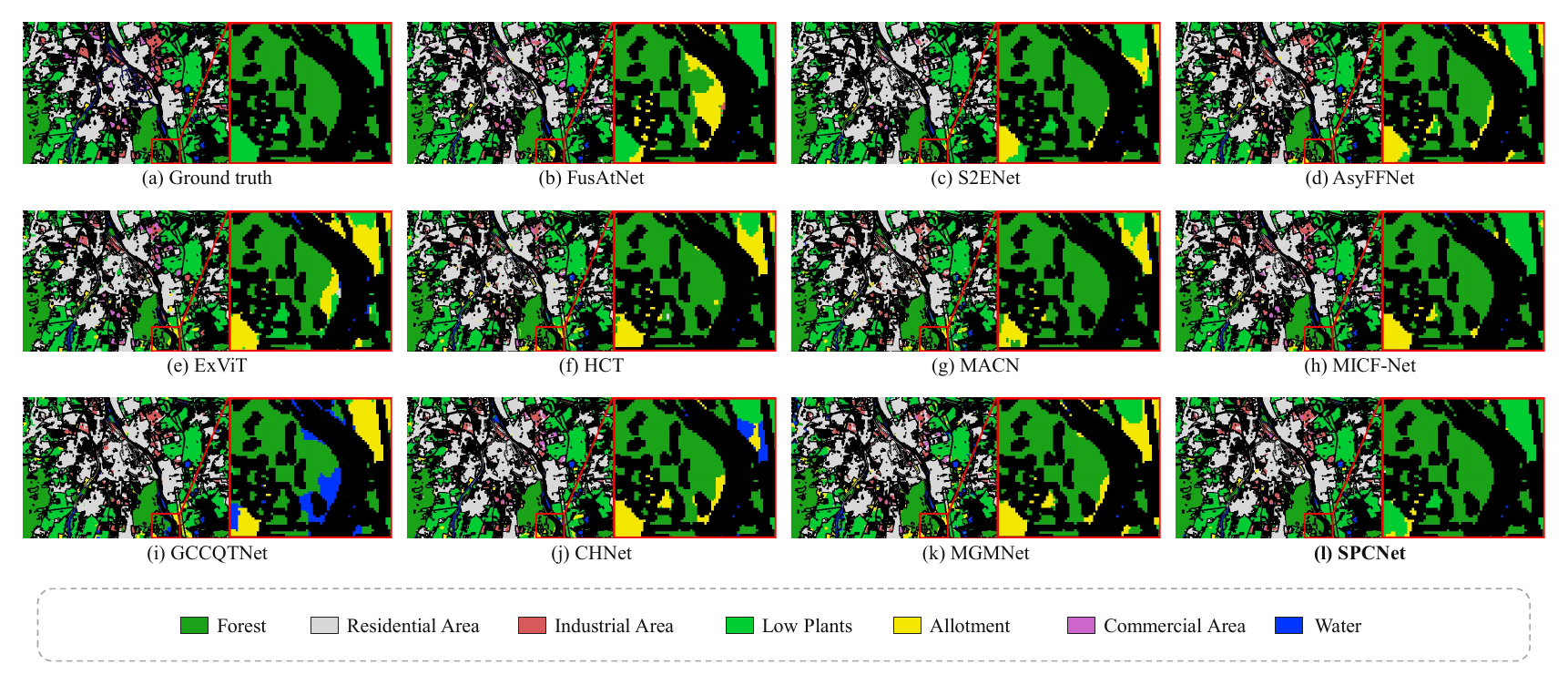}
\caption{Visualized classification results of different methods for the Augsburg dataset. (a) Ground truth. (b) FusAtNet. (c) S2ENet. (d) AsyFFNet. (e) ExViT. (f) HCT. (g) MACN. (h) MICF-Net. (i) GCCQTNet. (j) CHNet. (k) MGMNet. (l) SPCNet.}
\label{fig:augsvis}
\end{figure*}

\begin{figure*}
\centering
\includegraphics[width=0.98\textwidth]{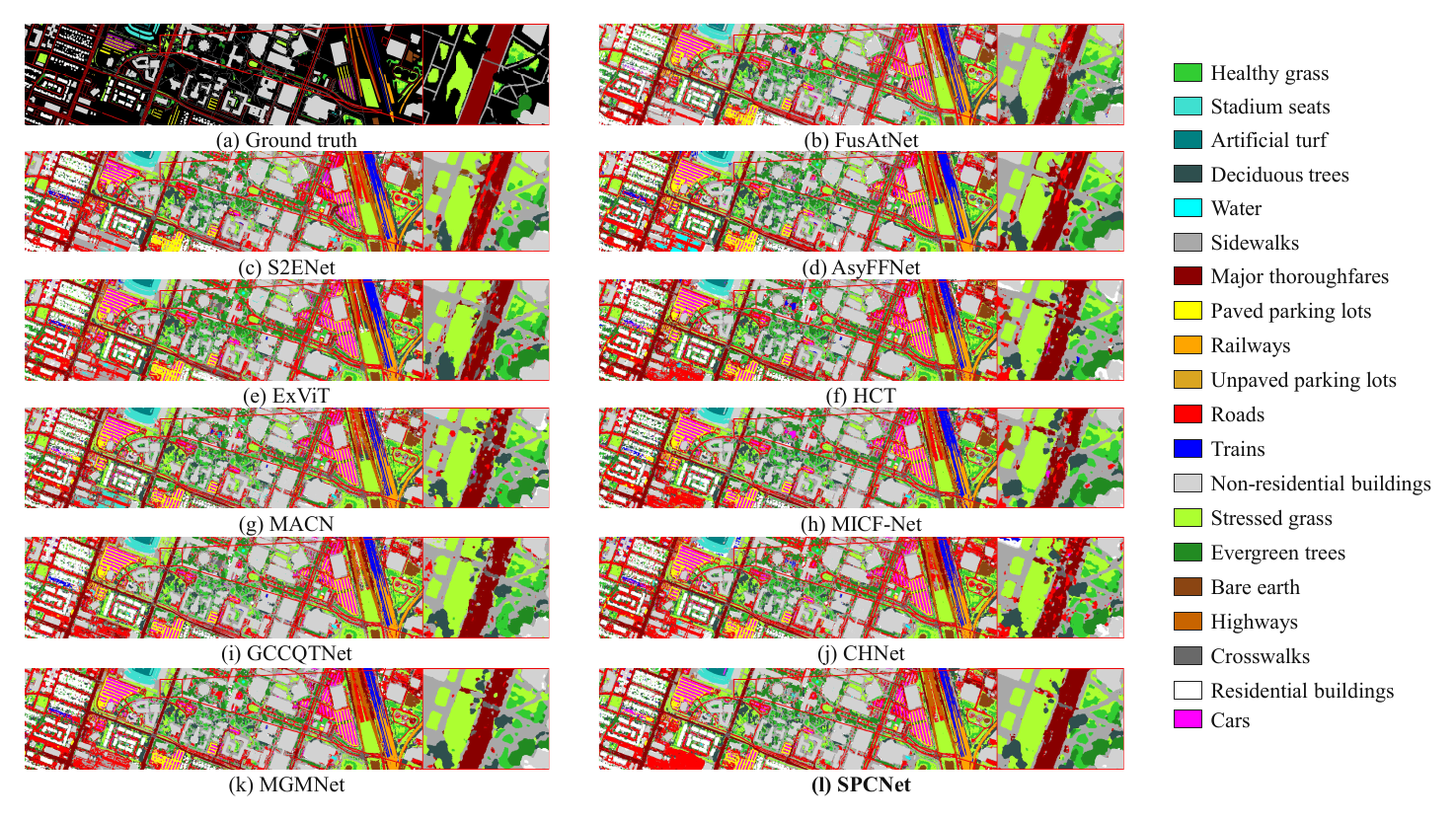}
\caption{Visualized classification results of different methods for the Houston2018 dataset. (a) Ground truth. (b) FusAtNet. (c) S2ENet. (d) AsyFFNet. (e) ExViT. (f) HCT. (g) MACN. (h) MICF-Net. (i) GCCQTNet. (j) CHNet. (k) MGMNet. (l) SPCNet.}
\label{fig:h18vis}
\end{figure*}

We compare SPCNet with ten representative state-of-the-art multi-source classification methods, following the comparison protocol commonly adopted in recent multi-source remote sensing studies. These baselines cover complementary design routes from attention-based CNNs to hybrid Transformers and high-order adaptive fusion. FusAtNet \cite{mohla2020fusatnet} applies self-attention to HSI spectral features and uses attention cues derived from LiDAR to enhance spatial representations. S2ENet \cite{fang2022s2enet} introduces a spatial-spectral cross-modal enhancement module to exchange information between HSI and LiDAR branches. AsyFFNet \cite{li2022asyffnet} adopts an asymmetric fusion strategy to account for the unequal information contributions of HSI and SAR. ExViT \cite{yao2023exvit} extends Vision Transformer with separable convolutions and a cross-modality attention module, while HCT \cite{zhao2022hct} combines hierarchical CNNs with a cross-token attention encoder for joint classification. MACN \cite{li2023macn} mixes self-attention and convolution so that local structures and global dependencies can be modeled in a unified layer. MICF-Net \cite{tang2024micf} employs dual-branch cross-source attention to extract global context and spatial relationship similarities, together with adaptive modulation to balance the optimization of different sources. GCCQTNet \cite{hu2025gccqtnet} uses an independent squeeze-expansion-like fusion structure and a cross-memory quaternion Transformer to reduce multimodal heterogeneity and model complex intra- and inter-modality relationships. CHNet \cite{ni2025chnet} progressively removes spectral-spatial redundancy and combines dynamic feature selection with factorized high-order learning to improve feature discriminability. MGMNet \cite{zhang2025mgmnet} embeds mutual guidance in a two-stream network, where context-aware filters generated from one source are used to refine the features of the other. These methods provide a representative coverage of local convolutional modeling, cross-attention, global token interaction, high-order fusion, and mutual-guidance strategies. Tables~\ref{tab:h13compare}--\ref{tab:h18compare} report the class-wise and overall performance, while Figs.~\ref{fig:h13vis}--\ref{fig:h18vis} show the corresponding classification maps. In addition to OA, AA, and Kappa, we examine confusing categories, boundary regions, and spatially fragmented objects to evaluate both numerical accuracy and spatial consistency.

\begin{table*}[!t]
\caption{Classification performance of different methods on the Houston2013 dataset. The bold entries denote the best results.}
\label{tab:h13compare}
\centering
\scriptsize
\setlength{\tabcolsep}{2pt}
\resizebox{0.90\textwidth}{!}{%
\begin{tabular}{c|ccccccccccc}
\hline\toprule
Class & FusAtNet & S2ENet & AsyFFNet & ExViT & HCT & MACN & MICF-Net & GCCQTNet & CHNet & MGMNet & \cellcolor{bg}SPCNet \\
\midrule
Healthy grass & 82.53 & 81.48 & 82.72 & 82.72 & 81.10 & 82.90 & 82.91 & 82.05 & \textbf{83.10} & 83.00 & \cellcolor{bg}82.91 \\
Stressed grass & 85.15 & 84.68 & 83.93 & 85.15 & 86.09 & \textbf{100.0} & 83.46 & 96.33 & 85.15 & 85.15 & \cellcolor{bg}99.72 \\
Synthetic grass & 98.81 & 99.01 & \textbf{100.0} & 99.21 & 99.80 & 99.80 & 99.80 & 99.41 & 99.60 & 99.60 & \cellcolor{bg}\textbf{100.0} \\
Trees & 92.90 & 92.14 & 91.95 & 91.95 & 98.01 & \textbf{98.11} & 97.63 & 91.00 & 92.61 & 87.59 & \cellcolor{bg}93.37 \\
Soil & \textbf{100.0} & \textbf{100.0} & \textbf{100.0} & \textbf{100.0} & \textbf{100.0} & \textbf{100.0} & \textbf{100.0} & 99.52 & 99.90 & \textbf{100.0} & \cellcolor{bg}\textbf{100.0} \\
Water & 98.60 & \textbf{100.0} & 95.80 & 98.60 & 95.10 & 95.80 & 95.80 & 95.80 & 95.80 & 89.51 & \cellcolor{bg}\textbf{100.0} \\
Residential & 85.45 & 89.93 & \textbf{96.18} & 86.66 & 83.49 & 78.17 & 83.49 & 90.95 & 88.25 & 95.62 & \cellcolor{bg}85.63 \\
Commercial & 81.81 & 90.18 & 81.14 & 91.05 & 86.43 & 80.46 & 81.91 & \textbf{95.67} & 81.62 & 76.42 & \cellcolor{bg}88.41 \\
Road & 83.76 & 92.63 & 86.12 & 91.12 & 86.69 & 90.93 & 82.15 & 87.54 & 89.61 & \textbf{93.96} & \cellcolor{bg}82.34 \\
Highway & 53.67 & 64.86 & 64.58 & 64.86 & 79.05 & 55.88 & 66.22 & 65.15 & 73.84 & 85.14 & \cellcolor{bg}\textbf{94.79} \\
Railway & 79.65 & 97.60 & 86.76 & 77.06 & 95.97 & 99.33 & 95.59 & 86.85 & 97.41 & 94.91 & \cellcolor{bg}\textbf{99.53} \\
Parking lot 1 & 91.26 & 88.09 & 90.59 & 88.76 & 98.46 & 99.71 & 94.62 & 99.14 & \textbf{99.90} & 98.37 & \cellcolor{bg}99.62 \\
Parking lot 2 & 81.05 & 91.93 & 92.28 & 87.72 & 90.88 & 88.77 & 81.40 & 82.81 & 89.47 & 89.12 & \cellcolor{bg}\textbf{96.49} \\
Tennis court & \textbf{100.0} & \textbf{100.0} & 95.14 & 97.17 & \textbf{100.0} & \textbf{100.0} & \textbf{100.0} & \textbf{100.0} & \textbf{100.0} & 99.60 & \cellcolor{bg}\textbf{100.0} \\
Running track & 98.73 & \textbf{100.0} & \textbf{100.0} & 98.52 & \textbf{100.0} & \textbf{100.0} & 99.79 & \textbf{100.0} & \textbf{100.0} & \textbf{100.0} & \cellcolor{bg}\textbf{100.0} \\
\midrule
OA & 85.32 & 89.56 & 87.95 & 87.42 & 90.66 & 89.81 & 88.09 & 90.40 & 90.30 & 90.98 & \cellcolor{bg}\textbf{93.53} \\
AA & 87.56 & 91.50 & 89.81 & 89.37 & 92.07 & 91.33 & 89.65 & 91.48 & 91.75 & 91.87 & \cellcolor{bg}\textbf{94.85} \\
Kappa & 84.12 & 88.69 & 86.96 & 86.38 & 89.86 & 88.93 & 87.12 & 89.59 & 89.47 & 90.20 & \cellcolor{bg}\textbf{92.98} \\
\bottomrule\hline
\end{tabular}}
\end{table*}

\textbf{Results on Houston2013 Dataset.} The quantitative results in Table~\ref{tab:h13compare} show that SPCNet achieves the best OA, AA, and Kappa values of 93.53\%, 94.85\%, and 92.98\%, respectively. Compared with the strongest competing OA of 90.98\% obtained by MGMNet, SPCNet provides an absolute gain of 2.55 percentage points. The improvement is particularly evident for difficult transportation-related classes: the accuracy for Highway reaches 94.79\%, exceeding the second-best result by 9.65 percentage points, while Railway and Parking lot 2 reach 99.53\% and 96.49\%, respectively. These results indicate that the shared large-kernel patterns capture extended spatial structures and that RAMA effectively exploits the complementary elevation information from LiDAR. As shown in Fig.~\ref{fig:h13vis}, several competing methods produce fragmented predictions and blurred boundaries in the enlarged region. SPCNet generates a map that is more consistent with the ground truth, with clearer object contours and fewer isolated errors. The agreement between the quantitative and visual results demonstrates strong discriminative capability in complex urban scenes.

\begin{table*}
\caption{Classification performance of different methods on the Augsburg dataset. The bold entries denote the best results.}
\label{tab:augscompare}
\centering
\scriptsize
\setlength{\tabcolsep}{2pt}
\resizebox{0.90\textwidth}{!}{
\begin{tabular}{c|ccccccccccc}
\hline\toprule
Class & FusAtNet & S2ENet & AsyFFNet & ExViT & HCT & MACN & MICF-Net & GCCQTNet & CHNet & MGMNet & \cellcolor{bg}SPCNet \\
\midrule
Forest & 93.78 & \textbf{98.10} & 96.01 & 90.04 & 96.92 & 97.95 & 95.98 & 95.26 & 96.24 & 96.28 & \cellcolor{bg}96.84 \\
Residential area & 97.58 & 99.08 & 97.96 & 95.44 & 98.96 & 98.68 & 98.46 & 96.80 & \textbf{99.22} & 97.76 & \cellcolor{bg}98.71 \\
Industrial area & 26.48 & 12.19 & 53.29 & 34.58 & 33.05 & 45.30 & 58.38 & \textbf{68.20} & 44.23 & 49.61 & \cellcolor{bg}64.10 \\
Low plants & \textbf{97.67} & 91.78 & 91.74 & 90.68 & 92.36 & 94.24 & 91.75 & 91.52 & 93.83 & 92.32 & \cellcolor{bg}97.08 \\
Allotment & 52.77 & 45.12 & \textbf{65.39} & 51.82 & 54.11 & 51.24 & 40.73 & 41.68 & 54.68 & 56.60 & \cellcolor{bg}52.39 \\
Commercial area & 24.66 & 1.22 & 4.58 & \textbf{28.63} & 6.62 & 6.74 & 23.53 & 13.97 & 19.85 & 5.15 & \cellcolor{bg}2.32 \\
Water & 47.51 & 24.09 & 49.64 & 17.65 & 47.40 & 49.27 & 44.81 & 46.80 & 46.74 & \textbf{49.93} & \cellcolor{bg}42.47 \\
\midrule
OA & 90.62 & 88.22 & 90.15 & 86.65 & 89.83 & 91.18 & 90.74 & 90.22 & 91.14 & 90.10 & \cellcolor{bg}\textbf{92.68} \\
AA & 62.92 & 53.08 & \textbf{65.51} & 58.40 & 61.34 & 63.35 & 64.81 & 64.89 & 64.97 & 63.95 & \cellcolor{bg}64.84 \\
Kappa & 86.33 & 86.47 & 85.74 & 80.79 & 85.28 & 87.23 & 86.65 & 86.00 & 87.10 & 85.75 & \cellcolor{bg}\textbf{89.37} \\
\bottomrule\hline
\end{tabular}}
\end{table*}

\textbf{Results on Augsburg Dataset.} Table~\ref{tab:augscompare} shows that SPCNet obtains the best OA and Kappa values of 92.68\% and 89.37\%, outperforming the second-best results by 1.50 and 2.14 percentage points, respectively. Its AA of 64.84\% remains competitive with the best value of 65.51\%. At the category level, SPCNet maintains strong results for Low plants, Forest, and Residential area, although the best score in each individual category is sometimes achieved by a competing method. The 2.32\% Commercial-area accuracy is consistent with severe class imbalance (only seven training pixels) and the coarse 30-m resolution of Augsburg's OpenStreetMap-derived functional labels \cite{hong21}, which may group mixed surfaces whose spectral and scattering responses overlap with Residential or Industrial areas. This sparse supervision and within-class variation depress minority-class recall and AA, while the common classes largely determine OA. The visual results in Fig.~\ref{fig:augsvis} further support this observation. SPCNet preserves more continuous forest and residential structures in the enlarged region and suppresses isolated errors that are visible in several competing maps. This demonstrates that collaborative HSI-SAR representation learning improves both regional consistency and robustness to scattering variations.

\begin{table*}
\caption{Classification performance of different methods on the Houston2018 dataset. The bold entries denote the best results.}
\label{tab:h18compare}
\centering
\scriptsize
\setlength{\tabcolsep}{2pt}
\resizebox{0.90\textwidth}{!}{
\begin{tabular}{c|ccccccccccc}
\hline\toprule
Class & FusAtNet & S2ENet & AsyFFNet & ExViT & HCT & MACN & MICF-Net & GCCQTNet & CHNet & MGMNet & \cellcolor{bg}SPCNet \\
\midrule
Healthy grass & 60.69 & \textbf{99.15} & 96.90 & 93.34 & 97.38 & 89.53 & 94.40 & 91.71 & 91.24 & 92.83 & \cellcolor{bg}92.26 \\
Stressed grass & 93.85 & 84.35 & 91.82 & 93.44 & 89.40 & 93.50 & 89.79 & 94.22 & 93.94 & 93.51 & \cellcolor{bg}\textbf{94.92} \\
Artificial turf & \textbf{100.0} & \textbf{100.0} & \textbf{100.0} & \textbf{100.0} & \textbf{100.0} & \textbf{100.0} & \textbf{100.0} & \textbf{100.0} & \textbf{100.0} & \textbf{100.0} & \cellcolor{bg}\textbf{100.0} \\
Evergreen trees & 99.02 & 98.13 & 99.29 & 95.87 & 98.90 & 99.10 & 99.24 & 98.58 & 98.50 & \textbf{99.63} & \cellcolor{bg}99.40 \\
Deciduous trees & 94.73 & 86.39 & \textbf{99.52} & 96.85 & 98.89 & 98.49 & 98.85 & 98.71 & 99.17 & 99.34 & \cellcolor{bg}99.04 \\
Bare earth & 99.87 & 99.33 & 99.98 & \textbf{100.0} & \textbf{100.0} & 99.96 & 99.84 & 99.96 & 99.98 & 99.96 & \cellcolor{bg}99.99 \\
Water & \textbf{100.0} & \textbf{100.0} & \textbf{100.0} & \textbf{100.0} & \textbf{100.0} & \textbf{100.0} & \textbf{100.0} & \textbf{100.0} & \textbf{100.0} & \textbf{100.0} & \cellcolor{bg}\textbf{100.0} \\
Residential buildings & 95.51 & 97.17 & 95.16 & \textbf{97.39} & 95.30 & 95.31 & 94.83 & 96.41 & 94.81 & 97.22 & \cellcolor{bg}97.13 \\
Non-residential buildings & 93.64 & 93.93 & 95.02 & 93.00 & 95.25 & 93.86 & 95.39 & 94.16 & 94.39 & 95.64 & \cellcolor{bg}\textbf{96.41} \\
Roads & 70.94 & 70.17 & 72.72 & 72.18 & 72.77 & 72.11 & 73.33 & 77.79 & \textbf{82.42} & 72.04 & \cellcolor{bg}78.13 \\
Sidewalks & 79.11 & 72.11 & 77.18 & 69.69 & 73.27 & 80.69 & 78.10 & 82.55 & 77.99 & 84.50 & \cellcolor{bg}\textbf{85.47} \\
Crosswalks & 86.06 & 82.47 & 93.14 & 88.12 & \textbf{98.95} & 97.69 & 97.63 & 97.11 & 97.61 & 97.98 & \cellcolor{bg}94.15 \\
Major thoroughfares & 76.12 & 89.44 & 84.67 & 86.12 & 80.90 & 79.05 & 77.73 & 88.03 & 83.84 & 89.64 & \cellcolor{bg}\textbf{90.96} \\
Highways & 98.89 & 98.53 & \textbf{99.53} & 99.50 & 98.01 & 96.86 & 97.04 & 98.43 & 96.80 & 99.41 & \cellcolor{bg}99.18 \\
Railways & 99.88 & 99.21 & \textbf{99.90} & 99.87 & 99.74 & \textbf{99.90} & 99.76 & 99.77 & 99.57 & \textbf{99.90} & \cellcolor{bg}99.76 \\
Paved parking lots & 95.88 & 95.62 & 99.18 & 97.02 & 97.07 & 97.74 & 98.72 & \textbf{99.35} & 97.22 & 97.77 & \cellcolor{bg}98.07 \\
Unpaved parking lots & \textbf{100.0} & \textbf{100.0} & \textbf{100.0} & \textbf{100.0} & \textbf{100.0} & \textbf{100.0} & \textbf{100.0} & \textbf{100.0} & \textbf{100.0} & \textbf{100.0} & \cellcolor{bg}\textbf{100.0} \\
Cars & 94.25 & 96.77 & 96.64 & 98.13 & 99.01 & 98.74 & 98.18 & 97.44 & 99.08 & \textbf{99.27} & \cellcolor{bg}98.97 \\
Trains & 99.84 & \textbf{100.0} & \textbf{100.0} & 99.98 & 99.97 & 99.74 & 99.88 & 99.89 & 99.97 & 99.99 & \cellcolor{bg}99.98 \\
Stadium seats & \textbf{100.0} & \textbf{100.0} & 99.89 & 99.99 & 99.98 & 99.74 & 99.98 & 99.84 & 99.95 & \textbf{100.0} & \cellcolor{bg}\textbf{100.0} \\
\midrule
OA & 89.09 & 90.05 & 91.24 & 89.98 & 90.56 & 90.31 & 90.65 & 92.13 & 91.75 & 92.63 & \cellcolor{bg}\textbf{93.77} \\
AA & 91.91 & 93.14 & 95.02 & 94.02 & 94.74 & 94.61 & 94.63 & 95.70 & 95.32 & 95.93 & \cellcolor{bg}\textbf{96.19} \\
Kappa & 85.96 & 87.20 & 88.71 & 87.14 & 87.81 & 87.55 & 87.94 & 89.86 & 89.37 & 90.48 & \cellcolor{bg}\textbf{91.93} \\
\bottomrule\hline
\end{tabular}}
\end{table*}

\textbf{Results on Houston2018 Dataset.} As reported in Table~\ref{tab:h18compare}, SPCNet achieves the best OA, AA, and Kappa values of 93.77\%, 96.19\%, and 91.93\%, respectively. Relative to the second-best results, the gains are 1.14 percentage points in OA, 0.26 percentage points in AA, and 1.45 percentage points in Kappa. SPCNet is especially effective for Stressed grass, Sidewalks, and Major thoroughfares, obtaining accuracies of 94.92\%, 85.47\%, and 90.96\%, respectively. The consistently high scores across vegetation, building, and transportation categories show that the proposed network balances spectral discrimination with structural modeling. In Fig.~\ref{fig:h18vis}, SPCNet produces cleaner class regions and more complete linear structures than the competing methods. Roads, railways, and the boundaries between vegetation and buildings are better retained in the enlarged region, while scattered prediction noise is reduced. This consistency between visual quality and quantitative performance confirms the generalization capability of SPCNet on a larger and more diverse urban scene.

\subsection{Ablation Study}

\textbf{Effectiveness of Essential Components.} To evaluate the contributions of SPCB and RAMA, Table~\ref{tab:ablation} compares the baseline, the two single-component variants, and the complete SPCNet using OA on all three datasets. The baseline obtains 88.56\%, 89.75\%, and 89.66\% on Houston2013, Augsburg, and Houston2018, respectively. The variant without SPCB, corresponding to the RAMA-only configuration, improves these results by 1.77, 1.67, and 2.09 percentage points. The variant without RAMA, corresponding to the SPCB-only configuration, improves OA by 2.36, 0.44, and 1.47 percentage points on Houston2013, Augsburg, and Houston2018, respectively. Thus, both modules improve the baseline, although their individual gains vary across datasets. When SPCB and RAMA are jointly incorporated, OA reaches 93.53\%, 92.68\%, and 93.77\%, which are 4.97, 2.93, and 4.11 percentage points above the baseline. The full model further exceeds the RAMA-only configuration by 3.20, 1.26, and 2.02 percentage points, and the SPCB-only configuration by 2.61, 2.49, and 2.64 percentage points. These results indicate that SPCB and RAMA provide complementary benefits under the same evaluation protocol.

\begin{table} 
\centering 
\caption{Ablation study of the proposed SPCNet in terms of OA (\%).} \label{tab:ablation} 
\begin{tabular}{lccc}
\toprule
Model Variant & \makecell{Houston\\2013} & Augsburg & \makecell{Houston\\2018} \\
\midrule
Baseline & 88.56 & 89.75 & 89.66 \\ 
w/o SPCB & 90.33 & 91.42 & 91.75 \\ 
w/o RAMA & 90.92 & 90.19 & 91.13 \\ 
\rowcolor{bg}Full SPCNet & \textbf{93.53} & \textbf{92.68} & \textbf{93.77} \\ 
\bottomrule
\end{tabular}
\end{table}

\begin{table}[!t]
\centering
\caption{Ablation study of SPCB in terms of OA (\%).}
\label{tab:featureblock}
\begin{tabular}{lccc}
\toprule
Model Variant & \makecell{Houston\\2013} & Augsburg & \makecell{Houston\\2018} \\
\midrule
Shared-LK only & 92.67 & 92.19 & 93.24 \\
DK only & 91.13 & 91.62 & 92.12 \\
\midrule
Static-DW (generator) & 93.21 & 91.78 & 93.33 \\
Static-DW (pointwise) & 92.66 & 91.90 & 93.26 \\
\midrule
\rowcolor{bg} Full SPCB & \textbf{93.53} & \textbf{92.68} & \textbf{93.77} \\
\bottomrule
\end{tabular}
\end{table}

\setcounter{topnumber}{3}
\begin{table}[!t]
\centering
\caption{Ablation study of RAMA in terms of OA (\%).}
\label{tab:rama}
\begin{tabular}{lccc}
\toprule
Fusion Strategy & \makecell{Houston\\2013} & Augsburg & \makecell{Houston\\2018} \\
\midrule
$1\times1$ convolution & 91.19 & 90.74 & 92.58 \\
Global allocator & \textbf{93.53} & 92.26 & 93.53 \\
Regional allocation & 93.29 & 92.31 & 93.63 \\
\rowcolor{bg} RAMA & \textbf{93.53} & \textbf{92.68} & \textbf{93.77} \\
\bottomrule
\end{tabular}
\end{table}

\begin{table}[!t]
\centering
\caption{Ablation study of the large-kernel sharing strategies in terms of OA (\%).}
\label{tab:lksharing}
\begin{tabular*}{0.84\columnwidth}{@{\extracolsep{\fill}}cc|ccc}
\toprule
\makecell{Modality\\Shared} & \makecell{Depth\\Shared} & \makecell{Houston\\2013} & Augsburg & \makecell{Houston\\2018} \\
\midrule
$\times$ & $\times$ & 90.46 & 91.42 & 92.28 \\
$\checkmark$ & $\times$ & 90.52 & 92.26 & 93.63 \\
$\times$ & $\checkmark$ & 91.34 & 91.76 & 93.33 \\
$\checkmark$ & $\checkmark$ & \textbf{93.53} & \textbf{92.68} & \textbf{93.77} \\
\bottomrule
\end{tabular*}
\end{table}

\textbf{Shared-Pattern Collaborative Block (SPCB).} Table~\ref{tab:featureblock} isolates the Shared Large-Kernel Pattern and Dynamic Kernel Generation while keeping the remaining network, including RAMA, unchanged. The Shared-LK-only and DK-only variants obtain 92.67\%, 92.19\%, and 93.24\%, and 91.13\%, 91.62\%, and 92.12\%, respectively, on Houston2013, Augsburg, and Houston2018. The Shared-LK-only variant is higher than the DK-only variant by 1.54, 0.57, and 1.12 percentage points, indicating a stronger individual contribution of the shared long-range pattern in these experiments. The complete SPCB reaches 93.53\%, 92.68\%, and 93.77\%, outperforming the Shared-LK-only variant by 0.86, 0.49, and 0.53 percentage points and the DK-only variant by 2.40, 1.06, and 1.65 percentage points. To assess input-conditioned filtering, Table~\ref{tab:featureblock} replaces the dynamic branch with two static controls of comparable capacity while retaining Shared-LK and the remaining network. Static-DW (generator) feeds a learned constant into the original generator; Static-DW (pointwise) uses a fixed depth-wise kernel with pointwise expansion and projection. Neither consistently improves on Shared-LK only: on Augsburg, OA drops by 0.41 and 0.29 percentage points, respectively, and both remain below Full SPCB on all three datasets. This may reflect the inability of fixed filters to adapt to input structures and their redundant responses with Shared-LK.

\begin{table*}[!t]
\centering
\caption{Comparison of model parameters, floating-point operations (FLOPs).}
\label{tab:cost}
\scriptsize
\setlength{\tabcolsep}{3pt}
\resizebox{0.92\textwidth}{!}{
\begin{tabular}{lccccccccccc}
\hline\toprule
Metrics & FusAtNet & S2ENet & AsyFFNet & ExViT & HCT & MACN & MICF-Net & GCCQTNet & CHNet & MGMNet & SPCNet \\
\midrule
Params (M) & 36.5760 & 1.1752 & 1.0654 & 0.2339 & 5.9186 & 2.3472 & 0.1332 & 1.2322 & 20.8156 & 4.3010 & 1.0893 \\
FLOPs (G) & 6.6950 & 0.2708 & 0.1747 & 0.0642 & 0.0849 & 0.3338 & 0.0377 & 0.0493 & 0.3214 & 0.1455 & 0.2470 \\
\bottomrule\hline
\end{tabular}}
\end{table*}

\textbf{Region-Adaptive Modality Allocator (RAMA).} Table~\ref{tab:rama} compares four fusion strategies with different levels of source-selection adaptivity. \textit{$1\times1$ convolution} concatenates the features and learns a linear channel mixture, but the same convolutional rule is applied at every spatial location. \textit{Global allocator} predicts one pair of modality-importance weights from a globally aggregated representation and applies the weights to the entire scene. \textit{Regional allocation} generates spatially varying allocation maps so that different regions can favor different sources; to isolate the contribution of regional allocation, its final fusion coefficients are fixed to $\alpha=\beta=0.5$. \textit{Full RAMA} combines the regional allocation maps with learnable fusion coefficients, enabling both spatial source selection and adaptive final aggregation.

The $1\times1$ convolution produces OA values of 91.19\%, 90.74\%, and 92.58\% on Houston2013, Augsburg, and Houston2018, respectively. The global allocator improves the results to 93.53\%, 92.26\%, and 93.53\%, demonstrating the benefit of learning source reliability, although one global weight pair cannot describe all local regions. Regional allocation obtains 93.29\%, 92.31\%, and 93.63\%, showing that spatially varying source selection is useful in Augsburg and Houston2018, although it is slightly lower than the global allocator on Houston2013. Full RAMA achieves 93.53\%, 92.68\%, and 93.77\%, tying the global allocator on Houston2013 and giving the best OA on the other two datasets. Compared with the global allocator, Full RAMA improves OA by 0, 0.42, and 0.24 percentage points, respectively; compared with regional allocation, it improves OA by 0.24, 0.37, and 0.14 percentage points. These results indicate that regional allocation and learnable final aggregation are complementary, yielding the best or tied-best OA across the three datasets.

\textbf{Shared-Pattern Collaborative Block (SPCB) Sharing Strategy.} Table~\ref{tab:lksharing} analyzes the effect of sharing the large-kernel operator across modalities and network depths.  Its first row corresponds to an independently parameterized setting in which the large-kernel operators are not shared across modalities or network depths, rather than a setting that removes the large-kernel branch. Without either form of sharing, OA is 90.46\%, 91.42\%, and 92.28\% on Houston2013, Augsburg, and Houston2018, respectively. Cross-modal sharing alone changes these results to 90.52\%, 92.26\%, and 93.63\%, giving gains of 0.06, 0.84, and 1.35 percentage points. Cross-depth sharing alone obtains 91.34\%, 91.76\%, and 93.33\%, improving OA by 0.88, 0.34, and 1.05 percentage points. When both sharing strategies are enabled, OA reaches 93.53\%, 92.68\%, and 93.77\%, exceeding the independently parameterized variant by 3.07, 1.26, and 1.49 percentage points. The combined strategy achieves the highest OA on all three datasets. These results support the complementarity of cross-modal and cross-depth sharing, while the dataset-dependent gains suggest that their relative contributions are not uniform across scenes. On Houston2013, the gain from cross-modal sharing increases from 0.06 to 2.19 percentage points when cross-depth reuse is enabled. Similar paved/built-up classes and narrow roads and railways make spatial structure informative \cite{houston2013}. Reusing the shared operator across modalities and depths helps retain this geometry as features evolve, while identity channels and dynamic filters preserve early and source-specific cues.

\subsection{Computational Complexity Analysis}

Table~\ref{tab:cost} compares the number of parameters and  floating-point operations (FLOPs) of different models. SPCNet contains 1.0893 M parameters and requires 0.2470 GFLOPs. Its parameter count is approximately 97.02\% and 94.77\% lower than those of FusAtNet and CHNet, respectively, indicating that the proposed modules do not rely on a heavyweight backbone. Although several highly lightweight competitors require less computation, SPCNet still has substantially lower FLOPs than FusAtNet, MACN, and CHNet. With controllable parameter and computational costs, the model achieves the highest overall accuracy (OA) on all three datasets. These results demonstrate that SPCNet achieves a favorable trade-off between accuracy and efficiency.

\subsection{Discussion}

Current experiments use registered multimodal imagery and do not quantify the effect of residual misregistration on cross-modal sharing. Future work can introduce controlled spatial offsets to measure this sensitivity and investigate registration-aware feature alignment to improve robustness.

RAMA learns location-wise weights under the classification objective to allocate branch contributions, but the correspondence between these weights and sensor quality remains to be assessed. Future work can introduce controlled modality-specific degradation to examine how allocations respond to changing input quality and develop quality-aware fusion. Although experiments cover three datasets and HSI--LiDAR and HSI--SAR combinations, they are evaluated using dataset-specific training and test splits, leaving their ability to generalize to unseen geographic scenes unexamined. Future work can combine cross-scene evaluation with domain generalization or lightweight adaptation to assess and improve transfer across acquisition conditions.

\section{Conclusion and Future Work}

We proposed SPCNet for multi-source remote sensing data classification. Specifically, the SPCB introduces a shared large-kernel mechanism and dynamic convolution to enhance common spatial-pattern learning while preserving modality-specific representations. Moreover, the RAMA adaptively assigns modality weights for different spatial regions, enabling more effective and reliable cross-modal feature fusion. Experimental results on the Houston2013, Augsburg, and Houston2018 datasets demonstrate that SPCNet achieves superior classification performance compared with several state-of-the-art methods. Future work will investigate more efficient dynamic-kernel generation and evaluate the proposed collaborative learning strategy in broader multi-source remote sensing scenarios.

\bibliographystyle{IEEEtran}
\bibliography{ref}

@article{gcz26tgrs,
  author={Gong, Chuanzheng and Gao, Feng and Lin, Junyan and Dong, Junyu and Du, Qian},
  journal={IEEE Transactions on Geoscience and Remote Sensing}, 
  title={Representative Spectral Correlation Network for Multisource Remote Sensing Image Classification}, 
  year={2026},
  volume={64},
  pages={1-14}}

@article{lv25ijde,
  title={Evaluating The potential of multi-source remote sensing and urban big data for urban land use classification},
  author={Lv, Qianwen and Qi, Zhixin and Li, Haibo and Song, Wenxuan and Huang, Di and Ding, Zihao and Shi, Qian},
  journal={International Journal of Digital Earth},
  volume={18},
  number={1},
  pages={2546519},
  year={2025}}

@article{sxy25ijdrr,
  title={A multiscale physically-based approach to urban flood risk assessment using ABM and multi-source remote sensing data},
  author={Shu, Xinyi and Ye, Chenlei and Xu, Zongxue and Liao, Ruting and Zhang, Silong},
  journal={International Journal of Disaster Risk Reduction},
  volume={119},
  pages={105332},
  year={2025}}

@article{hong26nrmp,
  title={Hyperspectral imaging},
  author={Hong, Danfeng and Li, Chenyu and Yokoya, Naoto and Zhang, Bing and Jia, Xiuping and Plaza, Antonio and Gamba, Paolo and Benediktsson, Jon Atli and Chanussot, Jocelyn},
  journal={Nature Reviews Methods Primers},
  volume={6},
  number={1},
  pages={19},
  year={2026}}

@article{ljy23ssmae,
  author={Lin, Junyan and Gao, Feng and Shi, Xiaochen and Dong, Junyu and Du, Qian},
  journal={IEEE Transactions on Geoscience and Remote Sensing}, 
  title={{SS-MAE}: Spatial--Spectral Masked Autoencoder for Multisource Remote Sensing Image Classification}, 
  year={2023},
  volume={61},
  pages={1-14},
  doi={10.1109/TGRS.2023.3331717}}

@article{xu2018tbcnn,
  author={Xu, Xiaodong and Li, Wei and Ran, Qiong and Du, Qian and Gao, Lianru and Zhang, Bing},
  journal={IEEE Transactions on Geoscience and Remote Sensing}, 
  title={Multisource Remote Sensing Data Classification Based on Convolutional Neural Network}, 
  year={2018},
  volume={56},
  number={2},
  pages={937-949},
  doi={10.1109/TGRS.2017.2756851}}

@article{fang2022s2enet,
  author={Fang, Sheng and Li, Kaiyu and Li, Zhe},
  journal={IEEE Geoscience and Remote Sensing Letters}, 
  title={{S2ENet}: Spatial--Spectral Cross-Modal Enhancement Network for Classification of Hyperspectral and {LiDAR} Data}, 
  year={2022},
  volume={19},
  pages={1-5},
  doi={10.1109/LGRS.2021.3121028}}

@inproceedings{mohla2020fusatnet,
  author={Mohla, Satyam and Pande, Shivam and Banerjee, Biplab and Chaudhuri, Subhasis},
  booktitle={2020 IEEE/CVF Conference on Computer Vision and Pattern Recognition Workshops (CVPRW)}, 
  title={{FusAtNet}: Dual Attention based SpectroSpatial Multimodal Fusion Network for Hyperspectral and {LiDAR} Classification}, 
  year={2020},
  pages={416-425}}

@article{gao2022dfinet,
  author={Gao, Yunhao and Li, Wei and Zhang, Mengmeng and Wang, Jianbu and Sun, Weiwei and Tao, Ran and Du, Qian},
  journal={IEEE Transactions on Geoscience and Remote Sensing}, 
  title={Hyperspectral and Multispectral Classification for Coastal Wetland Using Depthwise Feature Interaction Network}, 
  year={2022},
  volume={60},
  pages={1-15},
  doi={10.1109/TGRS.2021.3097093}}

@article{tang2024micf,
  author={Tang, Xu and Zou, Yizhou and Ma, Jingjing and Zhang, Xiangrong and Liu, Fang and Jiao, Licheng},
  journal={IEEE Transactions on Geoscience and Remote Sensing}, 
  title={Multiple Information Collaborative Fusion Network for Joint Classification of Hyperspectral and {LiDAR} Data}, 
  year={2024},
  volume={62},
  pages={1-16},
  doi={10.1109/TGRS.2024.3430373}}

@article{hu2025gccqtnet,
  author={Hu, Wen-Shuai and Li, Wei and Li, Heng-Chao and Huang, Feng-Hua and Tao, Ran},
  journal={IEEE Transactions on Neural Networks and Learning Systems}, 
  title={Global Clue-Guided Cross-Memory Quaternion Transformer Network for Multisource Remote Sensing Data Classification}, 
  year={2025},
  volume={36},
  number={4},
  pages={7357-7371},
  doi={10.1109/TNNLS.2024.3406735}}

@article{li2022asyffnet,
  author={Li, W. and Gao, Y. and Zhang, M. and Tao, R. and Du, Q.},
  journal={IEEE Transactions on Neural Networks and Learning Systems},
  title={Asymmetric Feature Fusion Network for Hyperspectral and {SAR} Image Classification},
  year={2023},
  volume={34},
  number={10},
  pages={8057--8070},
  doi={10.1109/TNNLS.2022.3149394}}

@article{yao2023exvit,
  author={Yao, J. and Zhang, B. and Li, C. and Hong, D. and Chanussot, J.},
  journal={IEEE Transactions on Geoscience and Remote Sensing},
  title={Extended Vision Transformer ({ExViT}) for Land Use and Land Cover Classification: A Multimodal Deep Learning Framework},
  year={2023},
  volume={61},
  pages={1--15},
  doi={10.1109/TGRS.2023.3284671}}

@article{zhao2022hct,
  author={Zhao, G. and Ye, Q. and Sun, L. and Wu, Z. and Pan, C. and Jeon, B.},
  journal={IEEE Transactions on Geoscience and Remote Sensing},
  title={Joint Classification of Hyperspectral and {LiDAR} Data Using a Hierarchical {CNN} and Transformer},
  year={2023},
  volume={61},
  pages={1--16},
  doi={10.1109/TGRS.2022.3232498}}

@article{li2023macn,
  author={Li, K. and Wang, D. and Wang, X. and Liu, G. and Wu, Z. and Wang, Q.},
  journal={IEEE Transactions on Geoscience and Remote Sensing},
  title={Mixing Self-Attention and Convolution: A Unified Framework for Multi-Source Remote Sensing Data Classification},
  year={2023},
  volume={61},
  pages={1--16},
  doi={10.1109/TGRS.2023.3310521}}

@article{ni2025chnet,
  author={Ni, K. and Xie, Y. and Zhao, G. and Zheng, Z. and Wang, P. and Lu, T.},
  journal={IEEE Transactions on Geoscience and Remote Sensing},
  title={Coarse-to-Fine High-Order Network for Hyperspectral and {LiDAR} Classification},
  year={2025},
  volume={63},
  pages={1--16},
  doi={10.1109/TGRS.2025.3554802}}

@article{zhang2025mgmnet,
  author={Zhang, W. and Song, W. and Wang, J. and Gao, W.},
  journal={IEEE Journal of Selected Topics in Applied Earth Observations and Remote Sensing},
  title={{MGMNet}: Mutual-Guidance Mechanism for Joint Classification of Multisource Remote Sensing Data},
  year={2025},
  volume={18},
  pages={1085--1097},
  doi={10.1109/JSTARS.2024.3502761}}

@article{hang2020coupledcnn,
  author={Hang, Renlong and Li, Zhu and Ghamisi, Pedram and Hong, Danfeng and Xia, Guiyu and Liu, Qingshan},
  journal={IEEE Transactions on Geoscience and Remote Sensing}, 
  title={Classification of Hyperspectral and {LiDAR} Data Using Coupled {CNNs}}, 
  year={2020},
  volume={58},
  number={7},
  pages={4939-4950},
  doi={10.1109/TGRS.2020.2969024}}

@article{l2022a3clnn,
  author={Li, Heng-Chao and Hu, Wen-Shuai and Li, Wei and Li, Jun and Du, Qian and Plaza, Antonio},
  journal={IEEE Transactions on Neural Networks and Learning Systems}, 
  title={{A3CLNN}: Spatial, Spectral and Multiscale Attention {ConvLSTM} Neural Network for Multisource Remote Sensing Data Classification}, 
  year={2022},
  volume={33},
  number={2},
  pages={747-761},
  doi={10.1109/TNNLS.2020.3028945}}

@article{zhang2023mtn,
  author={Zhang, Yiyan and Xu, Shufang and Hong, Danfeng and Gao, Hongmin and Zhang, Chenkai and Bi, Meiqiao and Li, Chenming},
  journal={IEEE Transactions on Geoscience and Remote Sensing}, 
  title={Multimodal Transformer Network for Hyperspectral and {LiDAR} Classification}, 
  year={2023},
  volume={61},
  pages={1-17},
  doi={10.1109/TGRS.2023.3283508}}

@article{wu2022multimodalcnn,
  author={Wu, Xin and Hong, Danfeng and Chanussot, Jocelyn},
  journal={IEEE Transactions on Geoscience and Remote Sensing}, 
  title={Convolutional Neural Networks for Multimodal Remote Sensing Data Classification}, 
  year={2022},
  volume={60},
  pages={1-10},
  doi={10.1109/TGRS.2021.3124913}}

@article{phz25jstars,
  author={Pan, Haizhu and Zhang, Quanxiu and Ge, Haimiao and Liu, Moqi and Shi, Cuiping},
  journal={IEEE Journal of Selected Topics in Applied Earth Observations and Remote Sensing}, 
  title={{PyionNet}: Pyramid Progressive Cross-Fusion Network for Joint Classification of Hyperspectral and {LiDAR} Data}, 
  year={2025},
  volume={18},
  pages={12042-12058},
  doi={10.1109/JSTARS.2025.3566214}}

@article{lc25tgrs,
  author={Liang, Chao and Zhao, Yuqing and Song, Yu and Ni, Kang},
  journal={IEEE Transactions on Geoscience and Remote Sensing}, 
  title={{IDNet}: Intensity-Constrained Detail-Enhanced Network for Hyperspectral and {LiDAR} Collaborative Classification}, 
  year={2025},
  volume={63},
  pages={1-15},
  doi={10.1109/TGRS.2025.3589143}}

@article{houston2013,
title={Hyperspectral and {LiDAR} Data Fusion: Outcome of the 2013 {GRSS} Data Fusion Contest},
author={Debes, C. and Merentitis, A. and Heremans, R. and Hahn, J. and Frangiadakis, N. and van Kasteren, T. and Liao, W. and Bellens, R. and Pizurica, A. and Gautama, S. and Philips, W. and Prasad, S. and Du, Q. and Pacifici, F.},
journal={IEEE Journal of Selected Topics in Applied Earth Observations and Remote Sensing},
volume={7},
number={6},
pages={2405--2418},
year={2014},
doi={10.1109/JSTARS.2014.2305441}}

@article{hong21,
title={Multimodal remote sensing benchmark datasets for land cover classification with a shared and specific feature learning model},
author={Hong, Danfeng and Hu, Jingliang and Yao, Jing and Chanussot, Jocelyn and Zhu, Xiao Xiang},
journal={ISPRS Journal of Photogrammetry and Remote Sensing},
volume={178},
pages={68--80},
year={2021},
doi={10.1016/j.isprsjprs.2021.05.011}}

@inproceedings{ding2022replknet,
  author={Ding, Xiaohan and Zhang, Xiangyu and Han, Jungong and Ding, Guiguang},
  title={Scaling Up Your Kernels to 31x31: Revisiting Large Kernel Design in {CNNs}},
  booktitle={Proceedings of the IEEE/CVF Conference on Computer Vision and Pattern Recognition},
  pages={11963--11975},
  year={2022},
  doi={10.1109/CVPR52688.2022.01166}}

@inproceedings{chen2020dynamicconv,
  author={Chen, Yinpeng and Dai, Xiyang and Liu, Mengchen and Chen, Dongdong and Yuan, Lu and Liu, Zicheng},
  title={Dynamic Convolution: Attention Over Convolution Kernels},
  booktitle={Proceedings of the IEEE/CVF Conference on Computer Vision and Pattern Recognition},
  pages={11030--11039},
  year={2020},
  doi={10.1109/CVPR42600.2020.01104}}

@inproceedings{yang2019condconv,
  author={Yang, Brandon and Bender, Gabriel and Le, Quoc V. and Ngiam, Jiquan},
  title={{CondConv}: Conditionally Parameterized Convolutions for Efficient Inference},
  booktitle={Advances in Neural Information Processing Systems},
  volume={32},
  year={2019}}

@article{li2019hsioverview,
  author={Li, Shutao and Song, Weiwei and Fang, Leyuan and Chen, Yushi and Ghamisi, Pedram and Benediktsson, Jon Atli},
  title={Deep Learning for Hyperspectral Image Classification: An Overview},
  journal={IEEE Transactions on Geoscience and Remote Sensing},
  year={2019},
  volume={57},
  number={9},
  pages={6690--6709},
  doi={10.1109/TGRS.2019.2907932}}

@article{rasti2020feature,
  author={Rasti, Behnood and Hong, Danfeng and Hang, Renlong and Ghamisi, Pedram and Kang, Xudong and Chanussot, Jocelyn and Benediktsson, Jon Atli},
  title={Feature Extraction for Hyperspectral Imagery: The Evolution From Shallow to Deep: Overview and Toolbox},
  journal={IEEE Geoscience and Remote Sensing Magazine},
  year={2020},
  volume={8},
  number={4},
  pages={60--88},
  doi={10.1109/MGRS.2020.2979764}}

@article{ghamisi2019fusionreview,
  author={Ghamisi, Pedram and Gloaguen, Richard and Atkinson, Peter M. and Benediktsson, Jon Atli and Rasti, Behnood and Yokoya, Naoto and Wang, Qunming and Hofle, Bernhard and Bruzzone, Lorenzo and Bovolo, Francesca and Chi, Mingmin and Anders, Katharina},
  title={Multisource and Multitemporal Data Fusion in Remote Sensing: A Comprehensive Review of the State of the Art},
  journal={IEEE Geoscience and Remote Sensing Magazine},
  year={2019},
  volume={7},
  number={1},
  pages={6--39},
  doi={10.1109/MGRS.2018.2890023}}

@article{hong2021diverse,
  author={Hong, Danfeng and Gao, Lianru and Yokoya, Naoto and Yao, Jing and Chanussot, Jocelyn and Du, Qian and Zhang, Bing},
  title={More Diverse Means Better: Multimodal Deep Learning Meets Remote-Sensing Imagery Classification},
  journal={IEEE Transactions on Geoscience and Remote Sensing},
  year={2021},
  volume={59},
  number={5},
  pages={4340--4354},
  doi={10.1109/TGRS.2020.3016820}}

@article{roy2023mft,
  author={Roy, Swalpa Kumar and Deria, Ankur and Hong, Danfeng and Rasti, Behnood and Plaza, Antonio and Chanussot, Jocelyn},
  title={Multimodal Fusion Transformer for Remote Sensing Image Classification},
  journal={IEEE Transactions on Geoscience and Remote Sensing},
  year={2023},
  volume={61},
  pages={1--20},
  doi={10.1109/TGRS.2023.3286826}}

@article{chang2024dsft,
  author={Chang, Honghao and Bi, Haixia and Li, Fan and Xu, Chen and Chanussot, Jocelyn and Hong, Danfeng},
  title={Deep Symmetric Fusion Transformer for Multimodal Remote Sensing Data Classification},
  journal={IEEE Transactions on Geoscience and Remote Sensing},
  year={2024},
  volume={62},
  pages={1--15},
  doi={10.1109/TGRS.2024.3476975}}

@article{zhao2022fgcn,
  author={Zhao, Xudong and Tao, Ran and Li, Wei and Philips, Wilfried and Liao, Wenzhi},
  title={Fractional Gabor Convolutional Network for Multisource Remote Sensing Data Classification},
  journal={IEEE Transactions on Geoscience and Remote Sensing},
  year={2022},
  volume={60},
  pages={1--18},
  doi={10.1109/TGRS.2021.3065507}}

@article{wang2023nncl,
  author={Wang, Meng and Gao, Feng and Dong, Junyu and Li, Heng-Chao and Du, Qian},
  title={Nearest Neighbor-Based Contrastive Learning for Hyperspectral and {LiDAR} Data Classification},
  journal={IEEE Transactions on Geoscience and Remote Sensing},
  year={2023},
  volume={61},
  pages={1--16},
  doi={10.1109/TGRS.2023.3236154}}

@article{gao2023csma,
  author={Gao, Yunhao and Zhang, Mengmeng and Wang, Junjie and Li, Wei},
  title={Cross-Scale Mixing Attention for Multisource Remote Sensing Data Fusion and Classification},
  journal={IEEE Transactions on Geoscience and Remote Sensing},
  year={2023},
  volume={61},
  pages={1--15},
  doi={10.1109/TGRS.2023.3263362}}

@article{zhang2023liit,
  author={Zhang, Yuwen and Peng, Yishu and Tu, Bing and Liu, Yaru},
  title={Local Information Interaction Transformer for Hyperspectral and {LiDAR} Data Classification},
  journal={IEEE Journal of Selected Topics in Applied Earth Observations and Remote Sensing},
  year={2023},
  volume={16},
  pages={1130--1143},
  doi={10.1109/JSTARS.2022.3232995}}

@article{sun2024mat,
  author={Sun, Le and Wang, Xinyu and Zheng, Yuhui and Wu, Zebin and Fu, Liyong},
  title={Multiscale 3-D--2-D Mixed {CNN} and Lightweight Attention-Free Transformer for Hyperspectral and {LiDAR} Classification},
  journal={IEEE Transactions on Geoscience and Remote Sensing},
  year={2024},
  volume={62},
  pages={1--16},
  doi={10.1109/TGRS.2024.3367374}}

@article{ni2024mhst,
  author={Ni, Kang and Wang, Duo and Zheng, Zhizhong and Wang, Peng},
  title={{MHST}: Multiscale Head Selection Transformer for Hyperspectral and {LiDAR} Classification},
  journal={IEEE Journal of Selected Topics in Applied Earth Observations and Remote Sensing},
  year={2024},
  volume={17},
  pages={5470--5483},
  doi={10.1109/JSTARS.2024.3366614}}

@article{ni2024fdbn,
  author={Ni, Kang and Wang, Duo and Zhao, Guofeng and Zheng, Zhizhong and Wang, Peng},
  title={Hyperspectral and {LiDAR} Classification via Frequency Domain-Based Network},
  journal={IEEE Transactions on Geoscience and Remote Sensing},
  year={2024},
  volume={62},
  pages={1--17},
  doi={10.1109/TGRS.2024.3435793}}

@article{han2023ssmunet,
  author={Han, Yi and Zhu, Hao and Jiao, Licheng and Yi, Xiaoyu and Li, Xiaotong and Hou, Biao and Ma, Wenping and Wang, Shuang},
  title={{SSMU-Net}: A Style Separation and Mode Unification Network for Multimodal Remote Sensing Image Classification},
  journal={IEEE Transactions on Geoscience and Remote Sensing},
  year={2023},
  volume={61},
  pages={1--15},
  doi={10.1109/TGRS.2023.3321729}}

@article{gao2025msfmamba,
  author={Gao, Feng and Jin, Xuepeng and Zhou, Xiaowei and Dong, Junyu and Du, Qian},
  title={{MSFMamba}: Multiscale Feature Fusion State Space Model for Multisource Remote Sensing Image Classification},
  journal={IEEE Transactions on Geoscience and Remote Sensing},
  year={2025},
  volume={63},
  pages={1--16},
  doi={10.1109/TGRS.2025.3535622}}

@article{song2023hdml,
  author={Song, Weiwei and Dai, Yong and Gao, Zhi and Fang, Leyuan and Zhang, Yongjun},
  title={Hashing-Based Deep Metric Learning for the Classification of Hyperspectral and {LiDAR} Data},
  journal={IEEE Transactions on Geoscience and Remote Sensing},
  year={2023},
  volume={61},
  pages={1--13},
  doi={10.1109/TGRS.2023.3321057}}

@article{ahmad2022hsisurvey,
  author={Ahmad, Muhammad and Shabbir, Sidrah and Roy, Swalpa Kumar and Hong, Danfeng and Wu, Xin and Yao, Jing and Khan, Adil Mehmood and Mazzara, Manuel and Distefano, Salvatore and Chanussot, Jocelyn},
  title={Hyperspectral Image Classification---Traditional to Deep Models: A Survey for Future Prospects},
  journal={IEEE Journal of Selected Topics in Applied Earth Observations and Remote Sensing},
  year={2022},
  volume={15},
  pages={968--999},
  doi={10.1109/JSTARS.2021.3133021}}

@article{hao2021garnn,
  author={Hao, Siyuan and Wang, Wei and Salzmann, Mathieu},
  title={Geometry-Aware Deep Recurrent Neural Networks for Hyperspectral Image Classification},
  journal={IEEE Transactions on Geoscience and Remote Sensing},
  year={2021},
  volume={59},
  number={3},
  pages={2448--2460},
  doi={10.1109/TGRS.2020.3005623}}

@article{roy2021a2s2kresnet,
  author={Roy, Swalpa Kumar and Manna, Suvojit and Song, Tiecheng and Bruzzone, Lorenzo},
  title={Attention-Based Adaptive Spectral--Spatial Kernel ResNet for Hyperspectral Image Classification},
  journal={IEEE Transactions on Geoscience and Remote Sensing},
  year={2021},
  volume={59},
  number={9},
  pages={7831--7843},
  doi={10.1109/TGRS.2020.3043267}}

@article{xue2022selfsupervised,
  author={Xue, Zhixiang and Yu, Xuchu and Yu, Anzhu and Liu, Bing and Zhang, Pengqiang and Wu, Shentong},
  title={Self-Supervised Feature Learning for Multimodal Remote Sensing Image Land Cover Classification},
  journal={IEEE Transactions on Geoscience and Remote Sensing},
  year={2022},
  volume={60},
  pages={1--15},
  doi={10.1109/TGRS.2022.3190466}}

@article{hu2023mdas,
  author={Hu, Jingliang and Liu, Rong and Hong, Danfeng and Camero, Andr{\'e}s and Yao, Jing and Schneider, Mathias and Kurz, Franz and Segl, Karl and Zhu, Xiao Xiang},
  title={{MDAS}: A New Multimodal Benchmark Dataset for Remote Sensing},
  journal={Earth System Science Data},
  year={2023},
  volume={15},
  number={1},
  pages={113--131},
  doi={10.5194/essd-15-113-2023}}

@article{ding2022glt,
  author={Ding, Kexing and Lu, Ting and Fu, Wei and Li, Shutao and Ma, Fuyan},
  title={Global--Local Transformer Network for {HSI} and {LiDAR} Data Joint Classification},
  journal={IEEE Transactions on Geoscience and Remote Sensing},
  year={2022},
  volume={60},
  pages={1--13},
  doi={10.1109/TGRS.2022.3216319}}

@article{li2025lsknet,
  author={Li, Yuxuan and Li, Xiang and Dai, Yimain and Hou, Qibin and Liu, Li and Liu, Yongxiang and Cheng, Ming-Ming and Yang, Jian},
  title={{LSKNet}: A Foundation Lightweight Backbone for Remote Sensing},
  journal={International Journal of Computer Vision},
  year={2025},
  volume={133},
  number={3},
  pages={1410--1431},
  doi={10.1007/s11263-024-02247-9}}

@misc{arevalo2017gmu,
  author={Arevalo, J. and Solorio, T. and Montes-y-G{\'o}mez, M. and Gonz{\'a}lez, F. A.},
  title={Gated Multimodal Units for Information Fusion},
  howpublished={arXiv preprint arXiv:1702.01992},
  year={2017}}

@article{gao2024ien,
  author={Gao, Hongmin and Feng, Hao and Zhang, Yiyan and Fei, Shuyu and Shen, Runhua and Xu, Shufang and Zhang, Bing},
  title={Interactive Enhanced Network Based on Multihead Self-Attention and Graph Convolution for Classification of Hyperspectral and {LiDAR} Data},
  journal={IEEE Transactions on Geoscience and Remote Sensing},
  year={2024},
  volume={62},
  pages={1--16}}

@article{gao26pys,
  author  = {Gao, Yinping and Li, Jiaxin and Wu, Yong and
             Zhang, Liping and Lei, Dajiang and Kong, Lingyan and
             Yue, Linfeng},
  title   = {{PyS}$^{2}${CF-Mamba}: A Pyramid Spatial--Spectral
             Competitive Fusion {Mamba} Network for
             Hyperspectral Image Classification},
  journal = {IEEE Geoscience and Remote Sensing Letters},
  year    = {2026},
  volume  = {23},
  pages   = {1--5}}

@article{li25edip,
  author  = {Li, Jiaxin and Zheng, Ke and Gao, Lianru and Han, Zhu
             and Li, Zhi and Chanussot, Jocelyn},
  title   = {Enhanced Deep Image Prior for Unsupervised Hyperspectral
             Image Super-Resolution},
  journal = {IEEE Transactions on Geoscience and Remote Sensing},
  year    = {2025},
  volume  = {63},
  pages   = {1--18}}

@article{li24m2unet,
  author  = {Li, Jiaxin and Zheng, Ke and Gao, Lianru and Ni, Li
             and Huang, Min and Chanussot, Jocelyn},
  title   = {Model-Informed Multistage Unsupervised Network for
             Hyperspectral Image Super-Resolution},
  journal = {IEEE Transactions on Geoscience and Remote Sensing},
  year    = {2024},
  volume  = {62},
  pages   = {1--17}}

\end{document}